\pdfoutput=1
\documentclass[11pt,twoside,a4paper,cmspaper,final,collab]{cms-tdr}
\def\svnVersion{20498ab}\def\svnDate{2022/04/21}\def\cmsCernNoTag{CERN-EP-2021-220}\def\cmsCernDate{\today}\def\cmsMessage{Published in the Journal of High Energy Physics as \href{http://dx.doi.org/10.1007/JHEP04(2022)087}{\doi{10.1007/JHEP04(2022)087}.}}
\begin{document}\cmsNoteHeader{B2G-20-013}

\newlength\cmsTabSkip\setlength{\cmsTabSkip}{1ex}
\providecommand{\cmsTable}[1]{\resizebox{\textwidth}{!}{#1}}

\newcommand{\Pj}{\ensuremath{{\mathrm{j}}}\xspace}
\newcommand{\PJ}{\ensuremath{{\mathrm{J}}}\xspace}

\newcommand{\V}{\PV}
\newcommand{\Ztoll}{\ensuremath{\PZ\to\Pell\Pell}\xspace}
\newcommand{\Ztott}{\ensuremath{\PZ\to\tau\tau}\xspace}
\newcommand{\llqq}{\ensuremath{2\Pell2\PQq}\xspace}
\newcommand{\ZV}{\ensuremath{\PZ\PV}\xspace}
\newcommand{\ZW}{\ensuremath{\PZ\PW}\xspace}
\newcommand{\ZZ}{\ensuremath{\PZ\PZ}\xspace}
\newcommand{\ZH}{\ensuremath{\PZ\PH}\xspace}
\newcommand{\ZX}{\ensuremath{\PZ\PX}\xspace}
\newcommand{\WW}{\ensuremath{\PW\PW}\xspace}
\newcommand{\Zjets}{\ensuremath{\PZ + \text{jets}}\xspace}
\newcommand{\mZV}{\ensuremath{m_{\ZV}}\xspace}
\newcommand{\mZH}{\ensuremath{m_{\ZH}}\xspace}
\newcommand{\mZX}{\ensuremath{m_{\ZX}}\xspace}
\newcommand{\mj}{\ensuremath{m_\mathrm{J}}\xspace}
\newcommand{\mjj}{\ensuremath{m_\mathrm{jj}}\xspace}
\newcommand{\mll}{\ensuremath{m_{\Pell\Pell}}\xspace}
\newcommand{\EM}{\ensuremath{\Pe\Pgm}\xspace}
\newcommand{\ee}{\ensuremath{\Pe\Pe}\xspace}
\newcommand{\tauo}{\ensuremath{\tau_{1}}\xspace}
\newcommand{\taut}{\ensuremath{\tau_{2}}\xspace}
\newcommand{\tauto}{\ensuremath{\tau_{21}}\xspace}
\newcommand{\B}{\ensuremath{\mathcal{B}}\xspace}
\newcommand{\Mplbar}{\ensuremath{\overline{M}_{\text{Pl}}}\xspace}
\newcommand{\ktilde}{\ensuremath{\widetilde{\kappa}}\xspace}
\newcommand{\DeltaR}{\ensuremath{\Delta R}\xspace}
\newcommand{\tplusX}{\ensuremath{\PQt + \text{X}}\xspace}
\DeclareRobustCommand{\alp}{{\HepParticle{a}{}{}}\xspace}
\DeclareRobustCommand{\astar}{{\HepParticle{a}{}{\ast}}\xspace}
\DeclareRobustCommand{\PSB}{{\HepSusyParticle{B}{}{}}\xspace}
\newcommand{\cgtil}{\ensuremath{c_\PXXSG}\xspace}
\newcommand{\cztil}{\ensuremath{c_\PSZ}\xspace}
\newcommand{\atwod}{\ensuremath{\widetilde{a}_{\text{2D}}}\xspace}
\newcommand{\falp}{\ensuremath{f_\alp}\xspace}

\cmsNoteHeader{B2G-20-013}
\title{Search for heavy resonances decaying to \texorpdfstring{\ZZ or \ZW}{ZZ or ZW} and axion-like particles mediating nonresonant \texorpdfstring{\ZZ or \ZH}{ZZ or ZH} production at \texorpdfstring{$\sqrt{s}=13\TeV$}{sqrt(s) = 13 TeV}}

\date{\today}

\abstract{
A search has been performed for heavy resonances decaying to \ZZ or \ZW and for axion-like particles (ALPs) mediating nonresonant \ZZ or \ZH production, in final states with two charged leptons ($\Pell=\Pe,\Pgm$) produced by the decay of a \PZ boson, and two quarks produced by the decay of a \PZ, \PW, or Higgs boson \PH. The analysis is sensitive to resonances with masses in the range 450 to 2000\GeV. Two categories are defined corresponding to the merged or resolved reconstruction of the hadronically decaying boson.   The search is based on data collected during 2016--2018 by the CMS experiment at the LHC in proton-proton collisions at a center-of-mass energy of 13\TeV, corresponding to an integrated luminosity of 138\fbinv. No significant excess is observed in the data above the standard model background expectation. Upper limits on the production cross section of heavy, narrow spin-2 and spin-1 resonances are derived as functions of the resonance mass, and exclusion limits on the production of bulk graviton particles and \PWpr bosons are calculated in the framework of the warped extra dimensions and heavy vector triplet models, respectively. In addition, upper limits on the ALP-mediated diboson production cross section and ALP couplings to standard model particles are obtained in the framework of linear and chiral effective field theories. These are the first limits on nonresonant ALP-mediated \ZZ and \ZH production obtained by the LHC experiments. 
}

\hypersetup{pdfauthor={CMS Collaboration},pdftitle={Search for heavy resonances decaying to ZZ or ZW and axion-like particles mediating nonresonant ZZ or ZH production at sqrt(s) = 13 TeV},pdfsubject={CMS},pdfkeywords={CMS, BSM, diboson, graviton, HVT, ALP}}

\maketitle 

\section{Introduction}\label{sec:introduction} 

Several extensions of the standard model (SM) predict the existence of new particles with large couplings to a pair of bosons, either Higgs bosons \PH or electroweak vector bosons $\PV = \PW, \PZ$.
One possibility is the Randall--Sundrum model with warped extra spatial dimensions (WED)~\cite{Agashe:2007zd,Randall:1999ee,Randall:1999vf}, which predicts heavy spin-2 gravitons (\PXXG).
Another example is the spin-1 heavy vector triplet (HVT) model~\cite{Pappadopulo:2014qza}, which postulates the existence of \PZpr and \PWpr bosons.

\begin{figure*}[htb]
  \centering
  \includegraphics[width=0.35\textwidth]{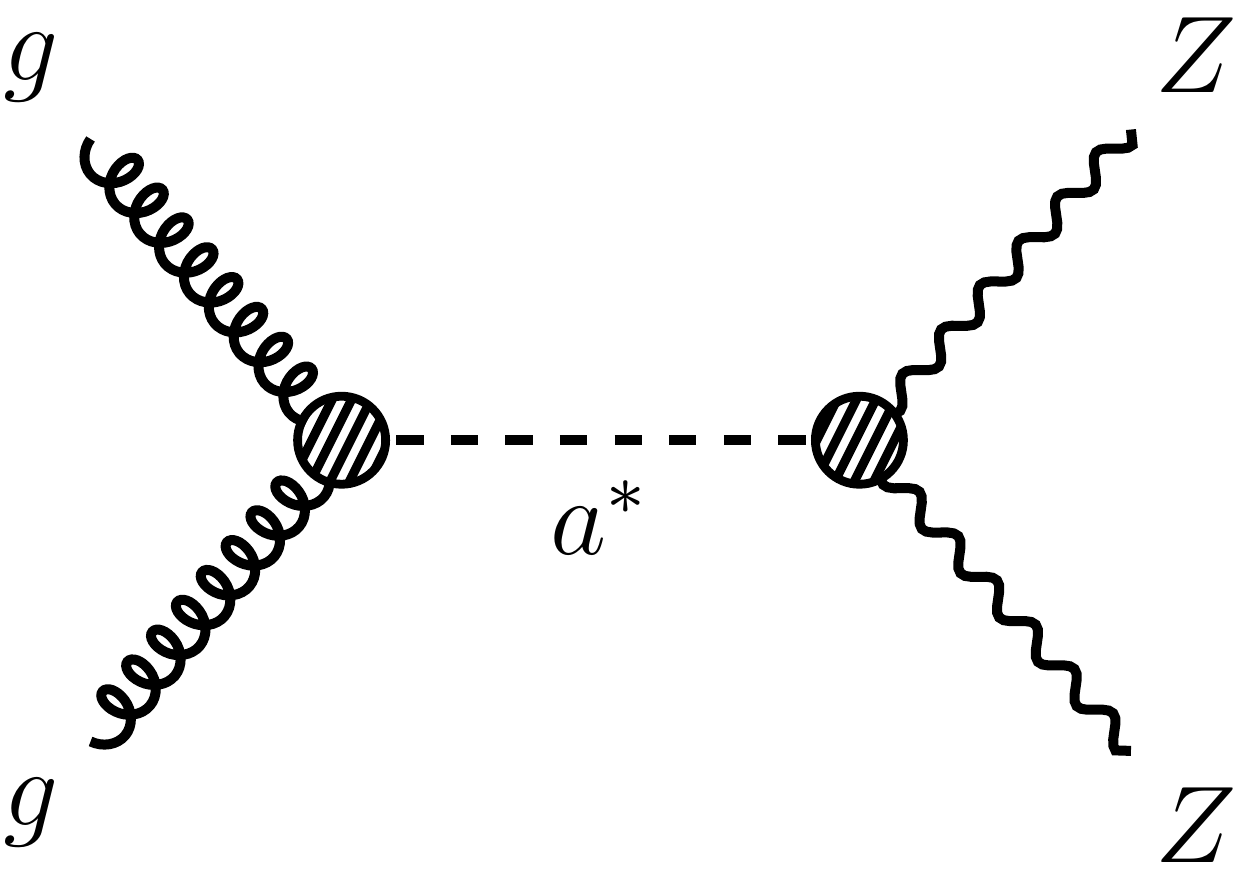}
  \hspace{2cm}
  \includegraphics[width=0.35\textwidth]{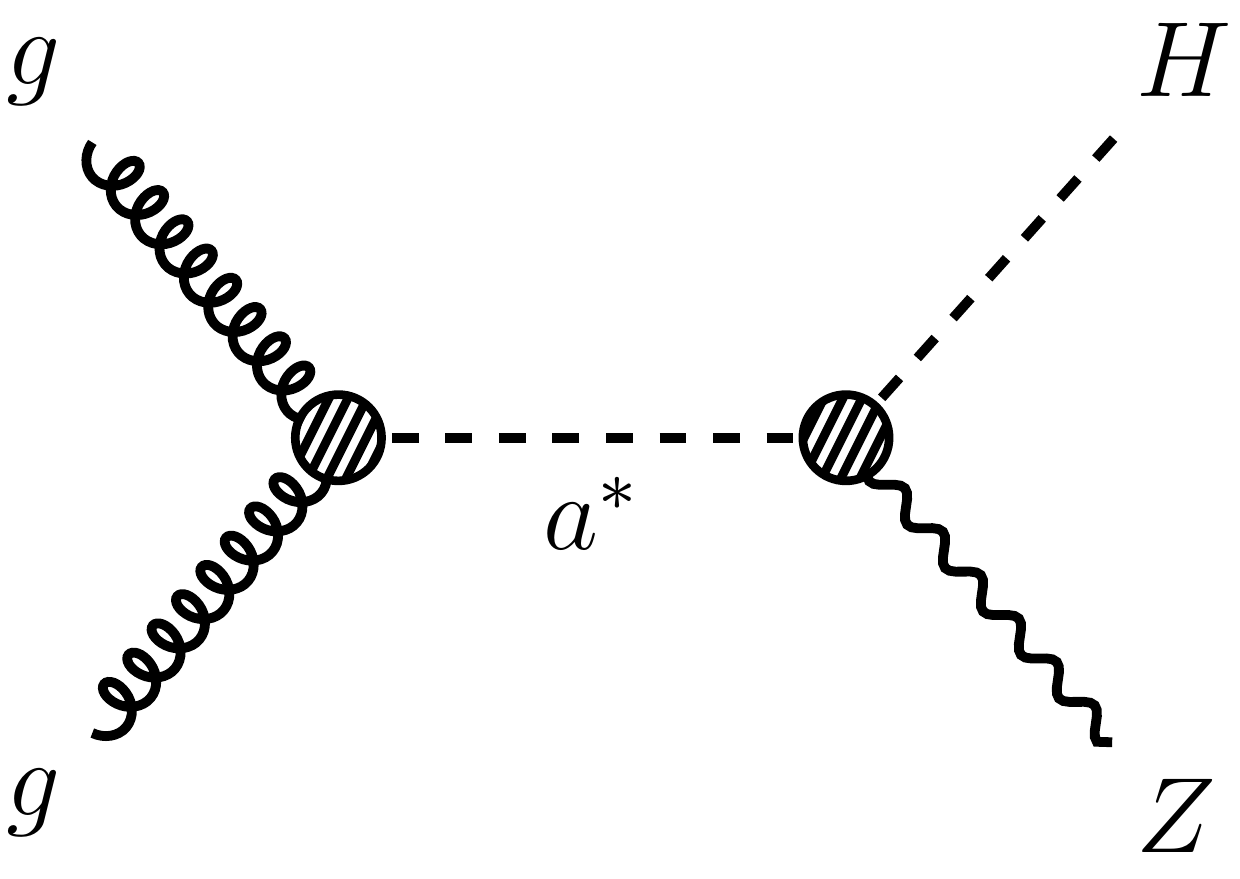}
  \caption{Feynman diagrams for the processes $\Pg\Pg \to \ZZ$ (left) and $\Pg\Pg \to \ZH$ (right)
    via an off-shell ALP \astar in the $s$ channel.}
  \label{fig:ALPFeynman}
\end{figure*}

Diboson signatures can also be used to search for axion-like particles (ALPs)~\cite{Georgi, Choi}, designated in this paper by ``\alp{}''. 
These are neutral pseudoscalar bosons with derivative interactions with SM particles, and  often appear in extensions of the SM.
Colliders allow for searches in a wide range of ALP masses and couplings~\cite{Brivio2, Bauer, NRALP}.
An interesting diboson production mechanism is nonresonant ALP-mediated scattering;
the ALP is an off-shell mediator in the $s$-channel $2 \to 2$ scattering processes~\cite{NRALP}.
Figure~\ref{fig:ALPFeynman} shows the Feynman diagrams for the processes $\Pg\Pg \to \astar \to \ZZ$ (left) and $\Pg\Pg \to \astar \to \ZH$ (right), where \astar is an off-shell ALP.
These channels are sensitive to the product of the ALP coupling to gluons and the relevant couplings
to dibosons. These channels have been identified as promising because the cross sections
are sufficiently large to enable strong constraints to be placed on the theoretical models using the 2016--2018 LHC data set.
The derivative interactions enhance the cross section at large diboson invariant masses, $\sqrt{\hat{s}}$.
In the ALP off-shell regime $\sqrt{\hat{s}} \gg m_\alp, m_\PZ$, where $m_\alp$ is the ALP mass,
the partonic cross section for $\Pg\Pg \to \astar \to \ZZ$ varies as $\hat{s} / f^4_\alp$, where \falp is the energy scale characterizing the new physics.
The same type of energy behavior holds for $\Pg\Pg \to \astar \to \ZH$.
Such energy dependence is only valid as long as the energies probed in the scattering process are smaller than \falp.
Factoring in the parton distribution functions (PDFs) of the proton, which tame the energy growth, the differential cross sections for the ALP-mediated processes $\Pp\Pp \to \ZZ, \ZH$ diminish more slowly as a function of $\sqrt{\hat{s}}$ than for the SM backgrounds.
The nonresonant cross sections and kinematical distributions
are found to be independent of the ALP mass from arbitrarily light masses up to masses of the order of 100\GeV~\cite{NRALP}.
The strategy of the search is to look for deviations with respect to the SM expectations in the tails of the experimental diboson mass distributions.

This paper reports on the results of a search for spin-2 gravitons decaying to \ZZ, spin-1 \PWpr bosons decaying to \ZW, and nonresonant ALP-mediated
\ZZ or \ZH production, in final states with two charged leptons ($\Pell=\Pe,\Pgm$) produced by the decay of a \PZ boson, and two quarks produced by the decay of a \PZ, \PW, or Higgs boson.
These two quarks can be reconstructed as one single ``merged'' jet in hadronic decays of a Lorentz-boosted boson, 
or as two individually resolved jets.
The analysis is sensitive to resonances with masses in the range from 450 to 2000\GeV.
The width of the heavy resonance is taken to be small in comparison to the experimental resolution~\cite{Khachatryan:2016kdb}.
This is the first time a search for the effects of nonresonant ALP-mediated \ZZ or \ZH production has been performed at the CERN LHC.
The search analyzes data collected from 2016 to 2018 by the CMS experiment at $\sqrt{s} = 13\TeV$.

Previous searches for heavy resonances decaying to two SM bosons at the LHC~\cite{Sirunyan:2018hsl,Sirunyan:2019vgt,CMS:2019qem,Sirunyan:2021bzu,CMS:2021klu,CMS:2021itu,ATLAS:2018sbw,ATLAS:2019nat,ATLAS:2020qiz,Aad:2020ddw,ATLAS:2020tlo} 
have found no evidence of a signal, providing stringent upper limits on the corresponding production cross sections.
Limits on ALP couplings derived from reinterpretations of ATLAS measurements can be found in Ref.~\cite{Carra:2021ycg}.

This paper is organized as follows: in Section~\ref{sec:datasim}, a description of the signal models, simulated background, and data samples used in the analysis is provided; Section~\ref{sec:detector} briefly describes the CMS detector; Section~\ref{sec:eventreconstruction} provides a description of the event reconstruction; in Section~\ref{sec:evtsel}, the event selection is discussed; Section~\ref{sec:bkg} describes the estimation of the SM background; the systematic uncertainties affecting the analysis are presented in Section~\ref{sec:sys}; and the results of the search for heavy spin-2 and spin-1 resonances and ALP-mediated diboson production are presented in Section~\ref{sec:res}. Finally, the results are summarized in Section~\ref{sec:conclusions}.

Tabulated results are provided in the HEPData record for this analysis~\cite{hepdata}.

\section{Data and simulated samples}\label{sec:datasim}

This analysis uses data collected with the CMS detector during proton-proton ($\Pp\Pp$) collisions at the LHC at $\sqrt{s}=13\TeV$, corresponding to an integrated luminosity of 138\fbinv. The events were selected online by criteria that require the presence of at least one electron or muon; these criteria are described in Section~\ref{sec:evtsel}.

Simulated signal samples are used in the analysis to optimize the search for production of heavy spin-2 and spin-1 resonances, and ALP-mediated \ZZ and \ZH production. Spin-2 and spin-1 resonant signal samples are generated according to the WED and HVT scenarios, respectively.

In the WED scenario~\cite{Fitzpatrick:2007qr,Goldberger:1999uk}, the main free parameters are the mass of the first Kaluza--Klein graviton excitation (the bulk graviton mass) and the ratio $\ktilde = \kappa/\Mplbar$, where $\kappa$ is a curvature parameter of the WED metric and \Mplbar is the reduced Planck mass. The coupling of the bulk graviton to vector bosons is determined by \ktilde.
Bulk graviton production cross sections, widths, and branching fractions are taken from Ref.~\cite{Oliveira:2014kla}.

The HVT model~\cite{Pappadopulo:2014qza} introduces a spin-1 triplet of \PZpr and \PWpr bosons. The HVT model generalizes a large number of explicit cases in terms of three parameters: $c_{\PH}$ determines the coupling to SM vector and Higgs bosons; $c_{\text{F}}$ determines the coupling to fermions; and $g_{\PV}$ is the overall strength of the new vector boson triplet interactions.
Benchmark ``Model A'', with $g_{\PV} = 1$, is representative of a model of weakly coupled vector resonances in an extension of the SM gauge group where the HVT bosons have comparable decay branching fractions into SM fermions and vector bosons.
Benchmark ``Model B'', with $g_{\PV} =3$, is representative of a composite model scenario where the HVT boson couplings to fermions are suppressed. 
The \PWpr boson production cross sections, widths, and branching fractions are taken from Ref.~\cite{Pappadopulo:2014qza}.
 
For both scenarios, the samples are generated at leading order (LO) in quantum chromodynamics (QCD) with the \MGvATNLO generator~\cite{MADGRAPH} versions 2.2.2 (2016) and 2.6.5 (2017 and 2018).
Different resonance mass hypotheses are considered in the range 450 to 2000\GeV.
Since the resonance width is small in comparison with the experimental resolution, for simplicity, the width is taken to be 1\MeV in the simulation (the narrow-width approximation). 
The generated spin-2 bulk graviton is forced to decay into two \PZ bosons, one decaying leptonically into any pair of charged leptons, and the other decaying hadronically into a pair of quarks.
In the case of the spin-1 \PWpr boson, the resonance is forced to decay into one \PZ and one \PW boson; additionally, the \PZ boson is then forced to decay to a pair of electrons, muons, or tau leptons, while the \PW boson is forced to decay into a pair of quarks.

The ALP interactions can be parameterized using the model-independent approach of effective field theories
(EFTs).
There are two EFT implementations: linear (related to weakly coupled new physics models,
with a minimal set of parameters)~\cite{Buchmuller, Grzadkowski} and chiral (related to strongly coupled
new physics models, with more parameters)~\cite{Feruglio,Azatov,Alonso2,Buchalla}. 
The ALP couplings to SM particles are proportional to a coefficient $c$ and inversely
proportional to the new physics energy scale \falp. Here, we adopt the definitions of Refs.~\cite{Brivio2, NRALP}.
Classical searches for ALPs at colliders consider their couplings to photons and gluons. The
coefficient of the gluon coupling is \cgtil. More recently, interest
has extended to consider ALP couplings to electroweak bosons: \ZZ, \WW, and $\PZ\gamma$~\cite{Brivio2}.
At LO in the linear model, all these and the coupling to photons are related by gauge symmetry
to two basic electroweak couplings: $c_\PSW$ and $c_\PSB$. The coupling of the ALP to \ZZ is
$\cztil = c_\PSW\ c^2_w + c_\PSB\ s^2_w$, where $c_w$ and $s_w$ denote the cosine and sine of the
Weinberg angle, respectively. In this analysis, small couplings of the ALP to fermions are set to zero.
In the chiral model, the ALP still couples to gluons with coupling \cgtil. Additionally, the ALP couples
to \ZH, and there are several coefficients contributing to the \ZH vertex~\cite{Brivio2}.
For simplicity, in this chiral analysis, the ALP to \ZH coupling coefficient \atwod is studied; small couplings of the ALP to fermions are neglected and the \ZZ channel is considered closed. In both the linear and chiral implementations, interference between the SM and ALP-mediated processes is neglected.

Simulated signals of ALP-mediated production are generated with \MGvATNLO 2.6.5 using both the linear and chiral EFT models~\cite{Brivio2}.
In the linear model, $\Pg\Pg \to \astar \to \ZZ$ events are generated at LO with 
$\cgtil/\falp = \cztil/\falp = 1\TeV^{-1}$ and an ALP mass of 1\MeV.
The \MGvATNLO cross sections are 77 (92, 94, 95)\unit{pb} for $\sqrt{\hat{s}} < \falp = 1$ (2, 3, 7)\TeV, respectively.
One \PZ boson is forced to decay leptonically and the other hadronically.
In the chiral model, $\Pg\Pg \to \astar \to \ZH$ events are generated at LO with
$\cgtil/\falp = \atwod/\falp = 1\TeV^{-1}$ and an ALP mass of 1\MeV; all other chiral coefficients are set to zero. 
The \MGvATNLO cross sections are 63 (78, 80, 81)\unit{pb} for $\sqrt{\hat{s}} < \falp = 1$ (2, 3, 7)\TeV, respectively.
The \PZ boson is forced to decay leptonically; the Higgs boson is decayed according to the branching fractions predicted in the SM.
In both cases, there is a 7\% systematic uncertainty in the size of the cross sections related to the renormalization and factorization scales, and a 1.2\% systematic uncertainty related to the PDFs.

Several SM processes yielding final states with charged leptons and jets are sources of background events for the analysis, and corresponding Monte Carlo simulated samples have been generated to study them.

The SM production of a \PZ boson in association with quarks or gluons in the final state (\Zjets) represents the dominant background process for the analysis, having topological similarities to the signal because of the presence of a pair of charged leptons and jets.
However, since the quark- and gluon-induced jets are not associated with the decay of a vector boson, the merged-jet and dijet mass spectra are characterized by smooth distributions, and the distribution of the \llqq system invariant mass falls exponentially. In contrast, the signals are expected to have peaking distributions in the merged-jet and dijet mass spectra, and a distinctive \llqq mass distribution.
The simulated \Zjets samples are produced with \MGvATNLO at next to LO (NLO) for 2016 samples, and LO for 2017 and 2018 samples. Both LO and NLO sample events are reweighted to account for electroweak NLO corrections, with the LO events also reweighted for QCD NLO corrections~\cite{Lindert:2017olm}.

The SM diboson production of \ZV and \ZH is an irreducible source of background for the analysis, since the merged-jet and dijet mass spectra will contain a peak from the hadronic decay of \PH, \PW, and \PZ bosons; however, this process produces a smoothly falling \llqq invariant mass distribution. The SM production of pairs of bosons (\ZZ, \ZW, and \ZH) is simulated at NLO with \MGvATNLO.

The parton showering and hadronization is simulated by interfacing the event generators with \PYTHIA~8.226~\cite{PYTHIA} with the CUETP8M1~\cite{CUETP8M1} tune for 2016 resonant samples and backgrounds, and \PYTHIA~8.230 with the CP5~\cite{CP5tune} tune for 2017 and 2018 resonant samples and backgrounds. The CUETP8M1 tune is used with \PYTHIA~8.230 for all ALP simulated samples. 
The PDFs of the colliding protons are given by the NNPDF~3.0~\cite{NNPDF30} PDF set for 2016 resonant samples and backgrounds, and the NNPDF~3.1~\cite{NNPDF31} set for 2017 and 2018 resonant samples and backgrounds, as well as all ALP simulated samples.
Additional $\Pp\Pp$ interactions occurring in the same or nearby bunch crossings (pileup) are added to the event simulation, with a frequency distribution adjusted to match that observed in data.
All samples are processed through a simulation of the CMS detector using \GEANTfour~\cite{GEANT4} and reconstructed using the same algorithms as those for analysis of the data.

\section{The CMS detector}\label{sec:detector}
The central feature of the CMS apparatus is a superconducting solenoid of 6\unit{m} internal diameter, providing a magnetic field of 3.8\unit{T}. 
Within the solenoid volume are a silicon pixel and strip tracker, a lead tungstate crystal electromagnetic calorimeter (ECAL), 
and a brass and scintillator hadron calorimeter (HCAL), each composed of a barrel and two endcap sections. 
Forward calorimeters extend the pseudorapidity $\eta$ coverage provided by the barrel and endcap detectors. 
Muons are measured in gaseous detectors embedded in the steel flux-return yoke outside the solenoid. 

Events of interest are selected using a two-tiered trigger system. The first level, composed of custom hardware processors, uses information from the calorimeters and muon detectors to select events at a rate of around 100\unit{kHz} within a fixed latency of about 4\mus~\cite{Sirunyan:2020zal}. The second level, known as the high-level trigger, consists of a farm of processors running a version of the full event reconstruction software optimized for fast processing, and reduces the event rate to around 1\unit{kHz} before data storage~\cite{Khachatryan:2016bia}.

A more detailed description of the CMS detector, together with a definition of the coordinate system used and the relevant kinematic variables, can be found in Ref.~\cite{Chatrchyan:2008zzk}.

\section{Event reconstruction}\label{sec:eventreconstruction}

The particle-flow (PF) algorithm~\cite{Sirunyan:2017ulk} aims to reconstruct and identify each individual particle in an event, with an optimized combination of information from the various elements of the CMS detector. The energy of photons is obtained from the ECAL measurement. The energy of electrons is determined from a combination of the electron momentum at the primary interaction vertex as determined by the tracker, the energy of the corresponding ECAL cluster, and the energy sum of all bremsstrahlung photons spatially compatible with originating from the electron track. The energy of muons is obtained from the curvature of the corresponding track. The energy of charged hadrons is determined from a combination of their momentum measured in the tracker and the matching ECAL and HCAL energy deposits, corrected for the response function of the calorimeters to hadronic showers. Finally, the energy of neutral hadrons is obtained from the corresponding corrected ECAL and HCAL energies. 

The candidate vertex with the largest value of summed physics-object $\pt^2$, where \pt is the transverse momentum, is taken to be the primary $\Pp\Pp$ interaction vertex. The physics objects are the jets, clustered using the jet finding algorithm~\cite{Cacciari:2008gp,Cacciari:2011ma} with the tracks assigned to candidate vertices as inputs, and the associated missing \pt, taken as the negative vector sum of the \pt of those jets.

The silicon tracker used in 2016 measured charged particles within the range $\abs{\eta} < 2.5$. For nonisolated particles of $1 < \pt < 10\GeV$ and $\abs{\eta} < 1.4$, the track resolutions are typically 1.5\% in \pt and 25--90 (45--150)\mum in the transverse (longitudinal) impact parameter~\cite{TRK-11-001}. At the start of 2017, the tracker was upgraded~\cite{Phase1Pixel} to measure up to $\abs{\eta} < 3.0$; the resolutions of the upgraded tracker were typically 1.5\% in \pt and 20--75\mum in the transverse impact parameter~\cite{DP-2020-049} for nonisolated particles of $1 < \pt < 10\GeV$ and $\abs{\eta} < 3.0$.

The electron momentum is estimated by combining the energy measurement in the ECAL with the momentum measurement in the tracker.
The momentum resolution for electrons with $\pt \approx 45\GeV$ from $\PZ \to \Pe\Pe$ decays ranges from 1.7 to 4.5\%. It is generally better in the barrel region than in the endcaps, and also depends on the bremsstrahlung energy emitted by the electron as it traverses the material in front of the ECAL~\cite{Khachatryan:2015hwa}.

Muons are measured in the range $\abs{\eta} < 2.4$, with detection planes made using three technologies: drift tubes, cathode strip chambers, and resistive-plate chambers. The single-muon trigger efficiency exceeds 90\% over the full $\eta$ range, and the efficiency to reconstruct and identify muons is greater than 96\%. Matching muons to tracks measured in the silicon tracker results in a relative \pt resolution, for muons with \pt up to 100\GeV, of 1\% in the barrel and 3\% in the endcaps, and of better than 7\% for muons in the barrel with \pt up to 1\TeV~\cite{Sirunyan:2018}. 

Both electrons and muons are required to be isolated from hadronic activity in the event.
An isolation variable is defined as the scalar sum of the \pt of charged hadrons originating from the primary vertex, plus the scalar sums of the \pt for neutral hadrons and photons, in a cone of $\DeltaR = \sqrt{\smash[b]{(\Delta\eta)^2 + (\Delta\phi)^2}} <0.3\,(0.4)$ around the electron (muon) direction, where $\phi$ is the azimuthal angle, corrected to account for the contribution from neutral candidates originating from pileup~\cite{Khachatryan:2015hwa,Sirunyan:2018}.

Hadronic jets are clustered from particles reconstructed by the PF algorithm using the infrared- and collinear-safe anti-\kt algorithm~\cite{Cacciari:2008gp,Cacciari:2011ma} with distance parameters of 0.4 (AK4 jets) and 0.8 (AK8 jets).
The jet momentum is determined as the vectorial sum of all particle momenta in the jet, and is found from simulation to be, on average, within 5--10\% of the true momentum over the entire \pt spectrum and detector acceptance. 
For AK4 jets, contamination from pileup is suppressed using charged-hadron subtraction, which removes from the list of PF candidates all charged particles originating from vertices other than the primary interaction vertex of the event. 
The residual contribution from neutral particles originating from pileup vertices is removed by means of an event-by-event jet-area-based correction to the jet four-momentum~\cite{Sirunyan:2020foa}.
For AK8 jets, the pileup per particle identification algorithm (PUPPI)~\cite{Bertolini:2014bba} is used to mitigate the effect of pileup at the reconstructed particle level, 
making use of local shape information, event pileup properties, and tracking information.

Jet energy corrections are derived from simulation studies so that the average measured energy of jets becomes identical to that of particle-level jets. In situ measurements of the momentum balance in dijet, $\text{photon} + \text{jet}$, $\PZ + \text{jet}$, and multijet events are used to determine any residual differences between the jet energy scale in data and in simulation, and appropriate corrections are made~\cite{Khachatryan:2016kdb}. Additional selection criteria are applied to each jet to remove jets potentially dominated by instrumental effects or reconstruction failures~\cite{CMS-PAS-JME-16-003}. 

A jet grooming technique is used for AK8 jets in this analysis to help identify and discriminate between merged jets from hadronic decays of boosted \PV and Higgs bosons, and jets from quarks and gluons.
The AK8 jets are groomed by means of the modified mass drop tagger algorithm~\cite{Dasgupta:2013ihk}, also known as the soft-drop algorithm, with angular exponent $\beta = 0$, soft cutoff threshold $z_{\text{cut}} < 0.1$, and characteristic radius $R_0 = 0.8$~\cite{Larkoski:2014wba}. Subjets are obtained by reversing the last step of the jet soft-drop clustering. The soft-drop algorithm does not fully reject contributions from the underlying event and pileup.
The mass of the AK8 jet (\mj) is therefore defined as the invariant mass associated to the four-momentum of the soft-drop jet, after the PUPPI corrections are applied.

Discrimination between AK8 jets originating from vector boson decays and those originating from gluons and quarks is further improved by the $N$-subjettiness jet substructure variable~\cite{Thaler2011}.
This observable exploits the distribution of the jet constituents found in the proximity of the subjet axes to determine if the jet can be effectively subdivided into a number $N$ of subjets.
The generic $N$-subjettiness variable $\tau_N$ is defined as the \pt-weighted sum of the angular distance of all the $k$ jet constituents from the closest subjet:
\begin{linenomath}
\begin{equation}
\tau_N = \frac{1}{d_0} \sum_k p_{\mathrm{T},k} \min( \Delta R_{1,k}, \Delta R_{2,k}, \dots, \Delta R_{N,k} ).
\end{equation}
\end{linenomath}
The normalization factor $d_0$ is defined as $d_0 = \sum_k p_{\mathrm{T},k} R_{\text{j}}$, with $R_{\text{j}}$ the clustering parameter of the original jet.
In this analysis, which aims to select $\V\to\PQq\PQq$ decays, the variable that best discriminates between \PV boson jets and those from quarks and gluons is the ratio of the 2-subjettiness to 1-subjettiness: $\tauto=\taut/\tauo$.
The \tauto observable is calculated for the jet before the grooming procedure, and includes the PUPPI algorithm corrections for pileup mitigation.

For the identification of jets originating from the hadronization of bottom quarks, the deep neural network-based combined secondary vertex (\textsc{DeepCSV}) algorithm~\cite{Sirunyan:2018btv} is used, either directly on the AK4 jets or on the AK8 soft-drop subjets with PUPPI pileup mitigation applied.
A \textsc{DeepCSV} loose (medium) \PQb tag corresponds to a $\approx$84 ($\approx$68)\% efficiency for \PQb quark identification and a $\approx$10 ($\approx$1)\% light-flavor or gluon jet mistag rate.
The \PQb tagging and mistag jet performances are extracted from simulation and then corrected using scale factors measured in data samples enriched in \PQb quark jets and light-flavor or gluon jets~\cite{Sirunyan:2018btv}.

Jets are required to be separated from isolated leptons by $\DR > 0.8\,(0.4)$ for AK8 (AK4) jets.
Only those AK4 and AK8 jets passing the tight jet identification requirements~\cite{CMS-PAS-JME-16-003}, and reconstructed in the central region of the detector ($\abs{\eta}<2.4$), are considered in the analysis.

\section{Event selection}\label{sec:evtsel}

In the online selection, events must satisfy either a single-electron or a single-muon trigger.
The electron trigger requires either an electron with $\pt > 27\,(32)\GeV$ and $\abs{\eta} < 2.4$ in the 2016 (2017 and 2018) data sample,
passing tight identification and isolation requirements,
or an electron with $\pt > 115\GeV$ and $\abs{\eta} < 2.4$  with no isolation requirements. 
The muon trigger requires a muon  with $\pt > 24\,(27)\GeV$ and $\abs{\eta}<2.4$ in 2016 and 2018 (2017), passing loose
identification and isolation requirements.
The muon trigger efficiency is larger than 99\% for the smallest resonance masses and decreases to about 96\% for the highest resonance masses.

In the offline selection, two well-identified leptons are first required.
The leading and subleading leptons are required to have $\pt > 40\GeV$
in the range $\abs{\eta}<2.4$. Electrons (muons) must satisfy tight identification
and tight (loose) isolation requirements.
In events with more than two leptons selected, the two with largest transverse momentum are considered.
Trigger, reconstruction, identification, and isolation efficiencies for leptons and their uncertainties 
are evaluated from dedicated data samples of leptonic \PZ boson decays, where one lepton from the decay
serves as a tag and the efficiency for the other lepton is measured~\cite{CMS:2011aa}.
A leptonic \PZ candidate is defined as a pair of leptons with the same flavor, opposite charges, and
invariant mass $76 < \mll < 106\GeV$.

Depending on the boson \pt in $\V\to\qqbar$ and $\PH\to\qqbar$ decays, 
the decay products may either be reconstructed as a single AK8 merged jet (boosted category) or as a pair of AK4 jets (resolved category). In the following, \Pj refers to AK4 jets and \PJ refers to AK8 jets.

\begin{figure*}[htbp]
\centering
\includegraphics[width=0.49\textwidth]{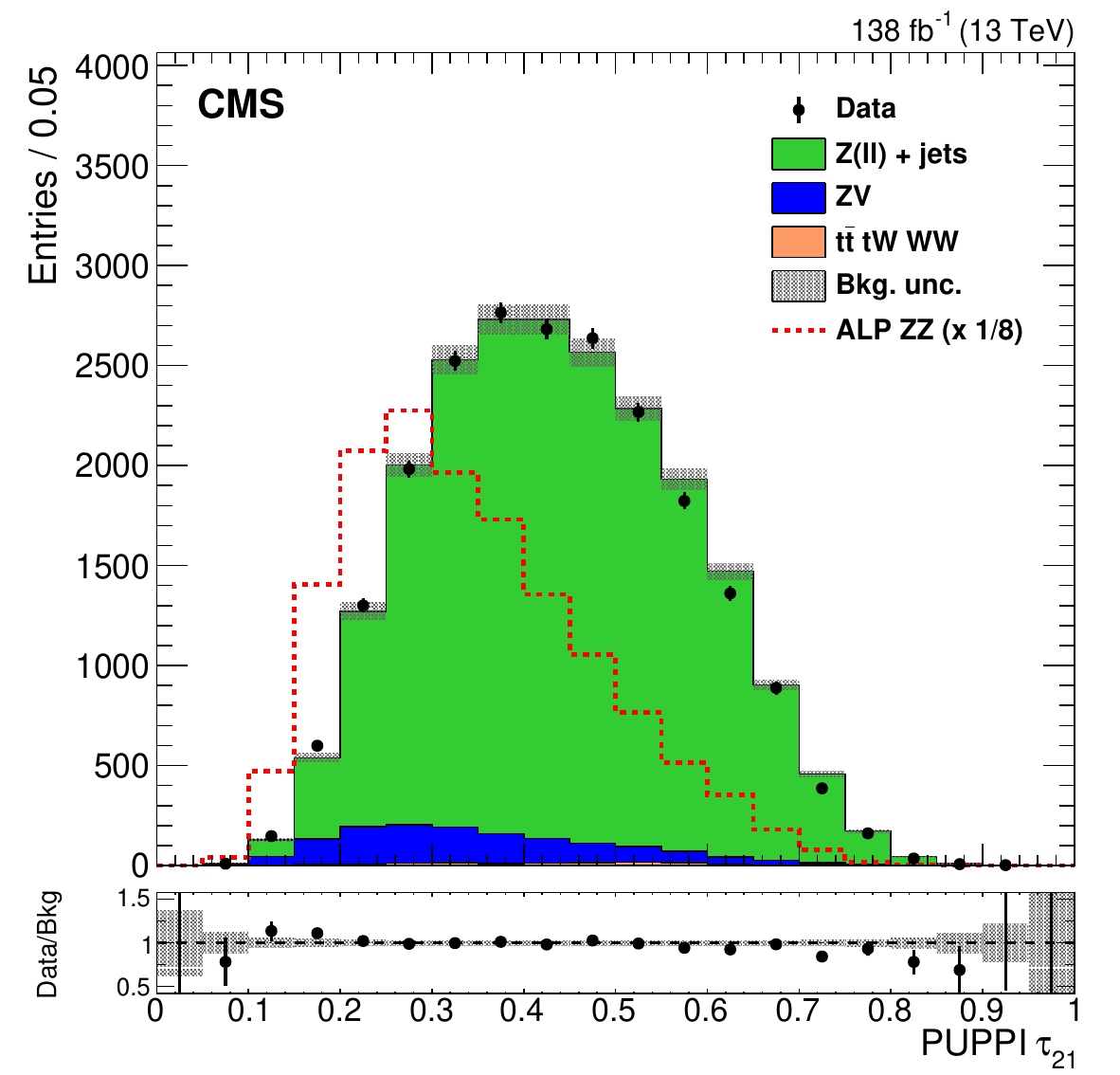}
\includegraphics[width=0.49\textwidth]{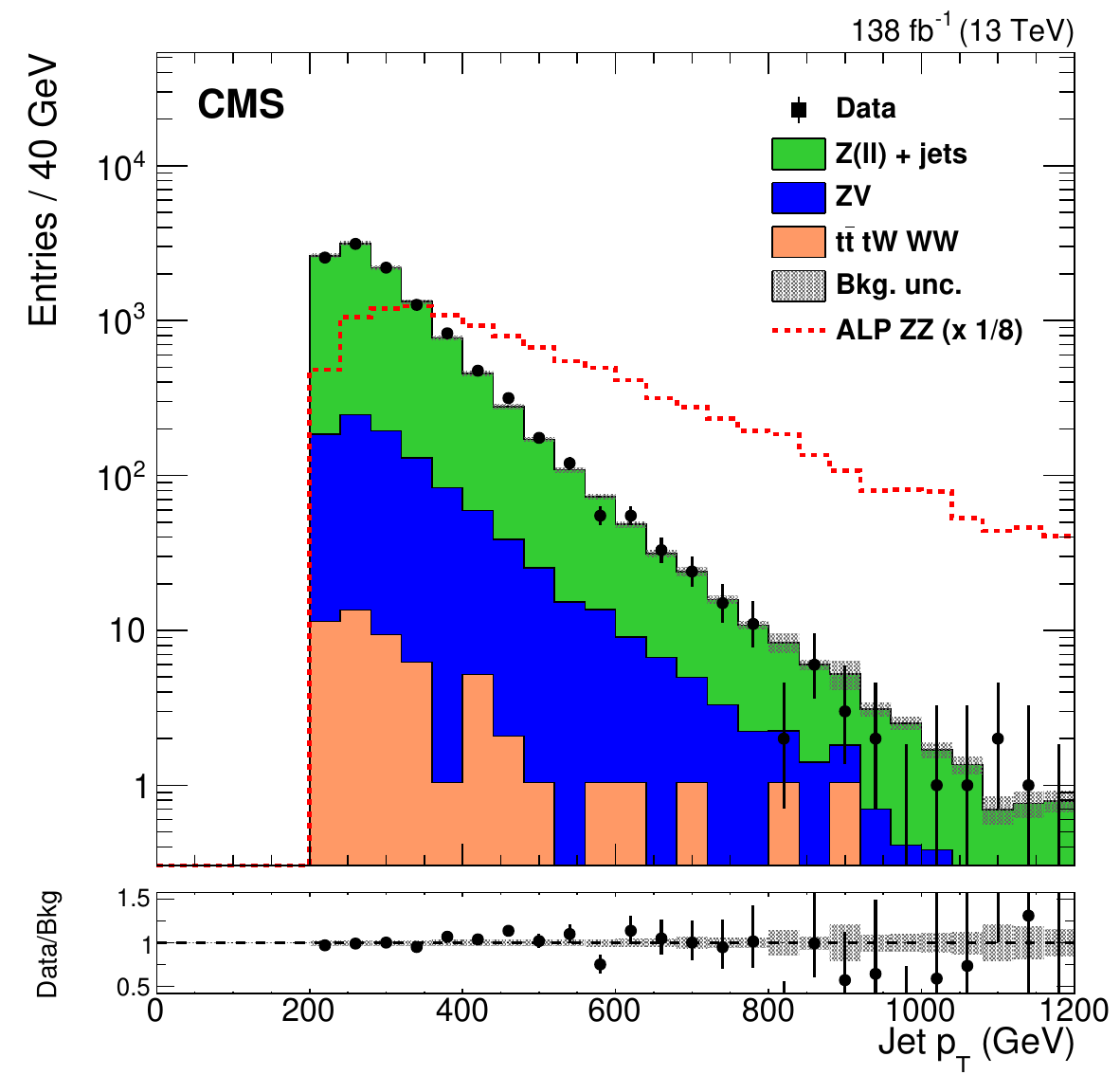}
\includegraphics[width=0.49\textwidth]{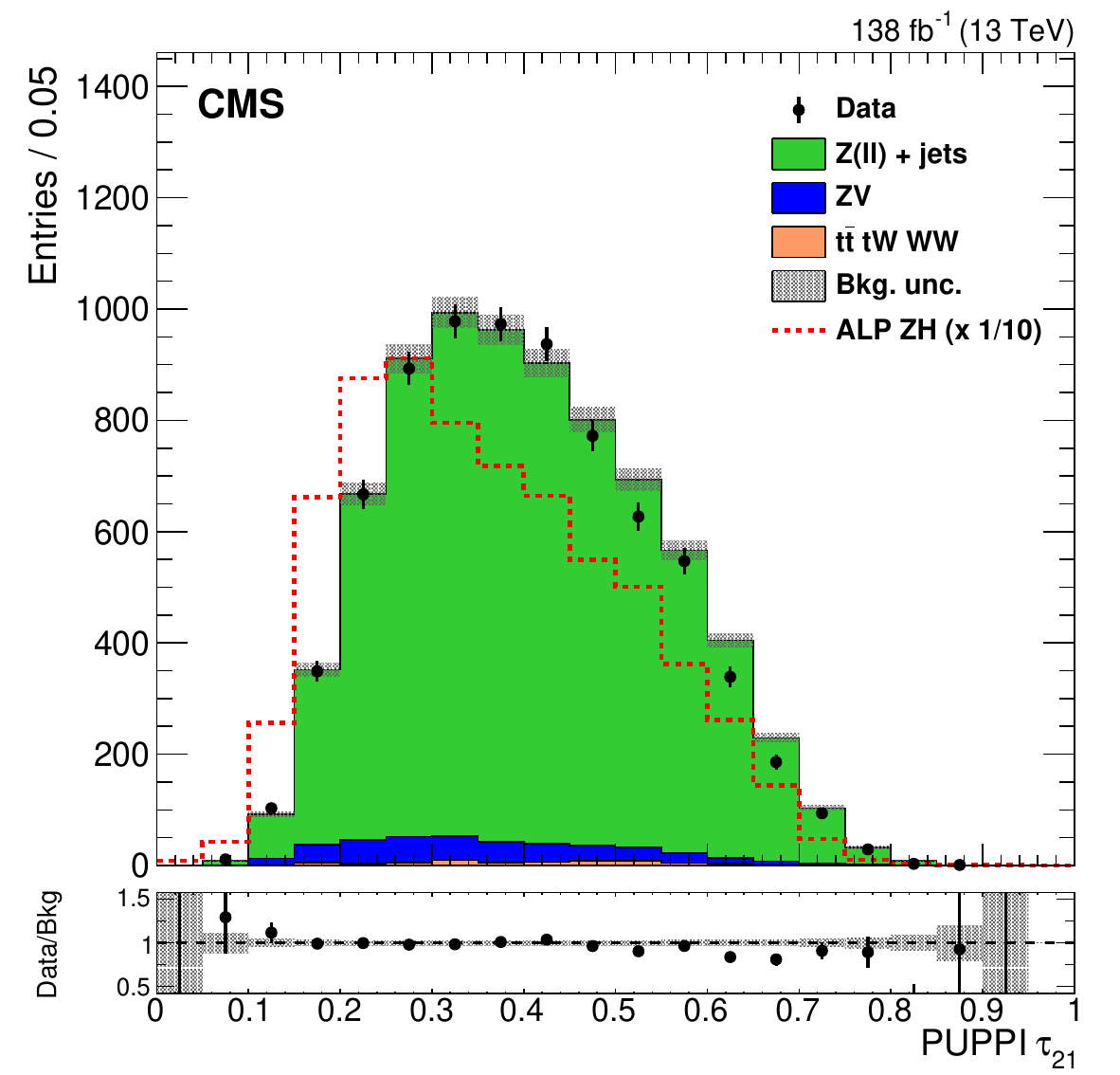}
\includegraphics[width=0.49\textwidth]{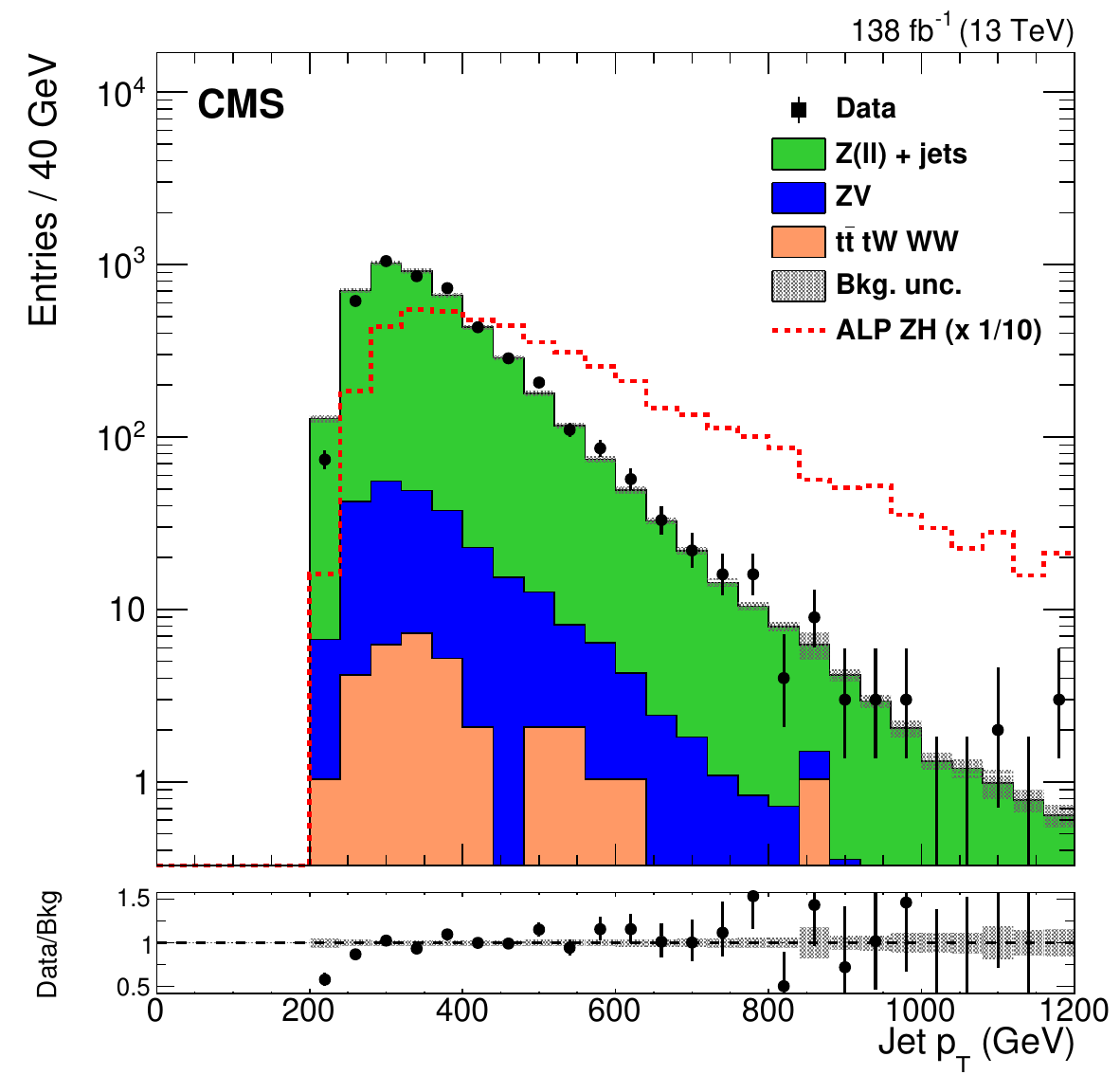}
\caption{Distributions of the merged jet \tauto (left) and \pt after applying the \tauto selection (right) for boosted hadronic \V (upper) and \PH (lower) candidates.
The gray band shows the statistical and systematic uncertainties in the background. 
The background for the \tauto distributions is normalized to the number of events in the data; the background normalization for the jet \pt distributions is derived from the final fit to the data.
The red dashed histograms correspond to a hypothetical linear (chiral) ALP with $1\TeV^{-1}$ couplings to gluons and \ZZ (\ZH), and $\falp = 3\TeV$; the cross sections have been multiplied by the factors indicated in the legends for better visibility. The lower panels show the ratio of data to background.
}
\label{fig:vvarsSRboosted}
\end{figure*}

In the boosted selection, events must have a leptonic \PZ candidate with $\pt(\Pell\Pell) > 200\GeV$ and one AK8 jet with $\pt(\PJ) > 200\GeV$ and soft-drop mass $\mj >30\GeV$. If several AK8 jets in the event satisfy the requirements, the one with the highest \pt is considered.
All AK8 jets have two subjets after the soft-drop mass requirement is satisfied.
Boosted hadronic \V and \PH candidates are first defined by $\tauto < 0.40\,(0.45)$ in 2016 (2017 and 2018), and then  
by the soft-drop mass signal regions $65 < \mj < 105\GeV$ (SR1) and $95 < \mj < 135\GeV$ (SR2), respectively.
Figure~\ref{fig:vvarsSRboosted} shows the \tauto and merged $\pt(\PJ)$ distributions for boosted hadronic \V and \PH candidates.
In addition, boosted hadronic sideband (SB) candidates are defined by the \tauto requirement and the soft-drop mass regions
$30 < \mj < 65\GeV$ and $135 < \mj < 300\GeV$.

In the resolved selection, boosted events are vetoed; events must have a leptonic \PZ candidate with $\pt(\Pell\Pell) > 150\GeV$ and an AK4 dijet system with  $\pt(\Pj) > 30\GeV$, $\pt(\Pj\Pj) > 150\GeV$, $\mjj > 30\GeV$, and $\DR(\Pj\Pj) < 1.5$.
In events with more than two AK4 jets, the best dijet combination is chosen based on the following criteria: 
dijets in the \PQb-tagged category are given priority; if more than one dijet combination remains, the pair closest
in mass to the target signal boson is chosen.
Resolved hadronic \V and \PH candidates are defined
by the dijet mass signal regions $65 < \mjj < 110\GeV$ (SR1) and $95 < \mjj < 135\GeV$ (SR2), respectively.
In addition, resolved hadronic SB candidates are defined by the dijet mass regions
$30 < \mjj < 65\GeV$ and $135 < \mjj < 180\GeV$.

\begin{figure*}[htbp]
\centering
\includegraphics[width=0.49\textwidth]{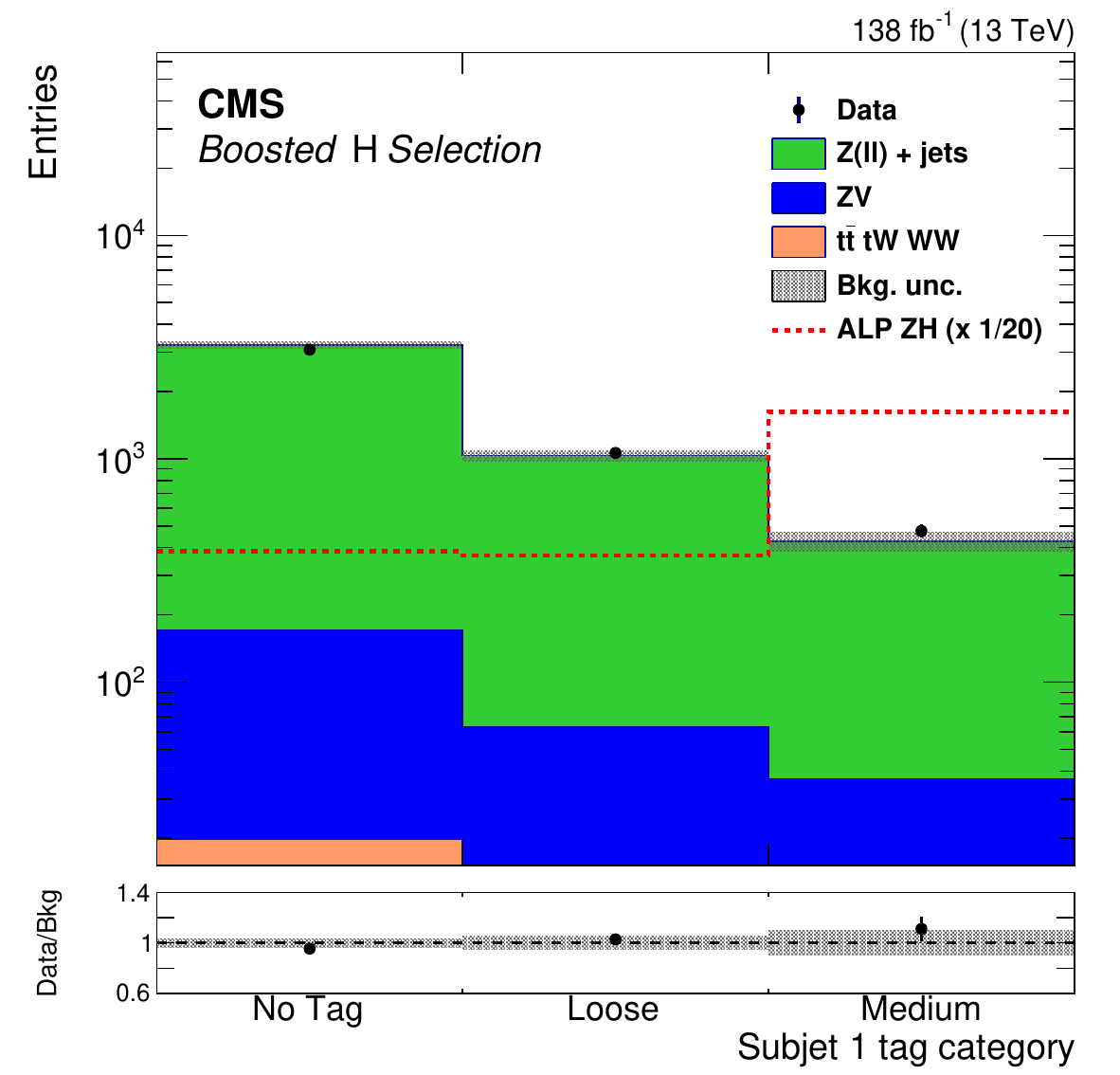}
\includegraphics[width=0.49\textwidth]{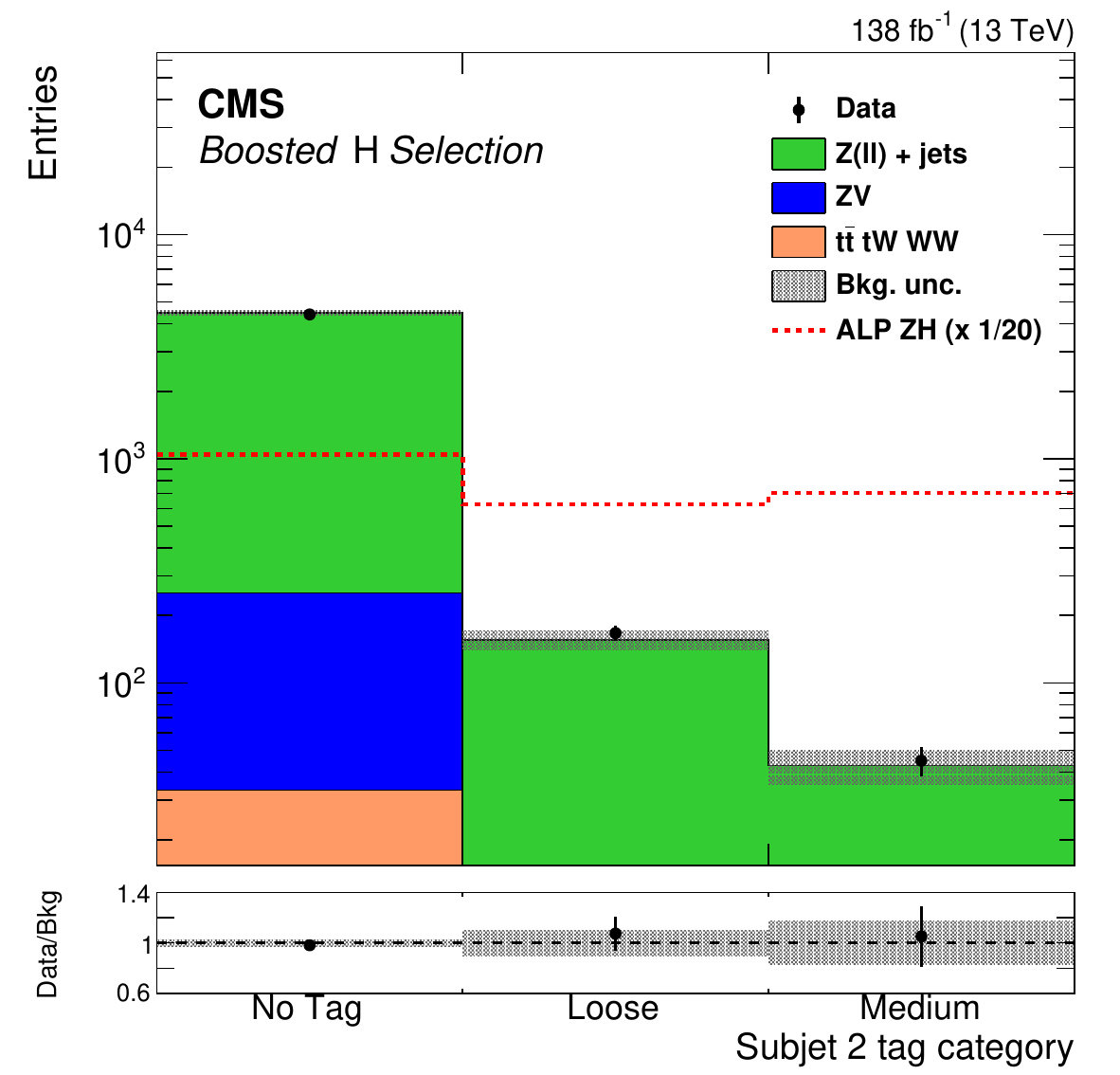}
\caption{
Distributions of the untagged, exclusive loose, and medium \textsc{DeepCSV} tags for the more \PQb-like subjet (left) and the less \PQb-like subjet (right) of the boosted hadronic \PH candidates in SR2. 
The gray band shows the statistical and systematic uncertainties in the background.
Background normalizations are derived from the final fit to the data.
The red dashed histograms correspond to a hypothetical chiral ALP with $1\TeV^{-1}$ couplings to gluons and \ZH, and $\falp = 3\TeV$; the cross sections have been multiplied by the factors indicated in the legends for better visibility. The lower panels show the ratio of data to background.
} 
\label{fig:btagCtg}
\end{figure*}

Boosted (resolved) events belong to the tagged category if the hadronic \V or \PH candidate contains one subjet (jet)
with a medium \PQb tag and the other subjet (jet) with a loose \PQb tag. Otherwise, they are placed in the
untagged category.
Figure~\ref{fig:btagCtg} shows the distributions of the untagged, exclusive loose, and medium \textsc{DeepCSV} tags for subjets of the boosted hadronic \PH candidates in SR2.

\begin{figure*}[htbp]
\centering
\includegraphics[width=0.49\textwidth]{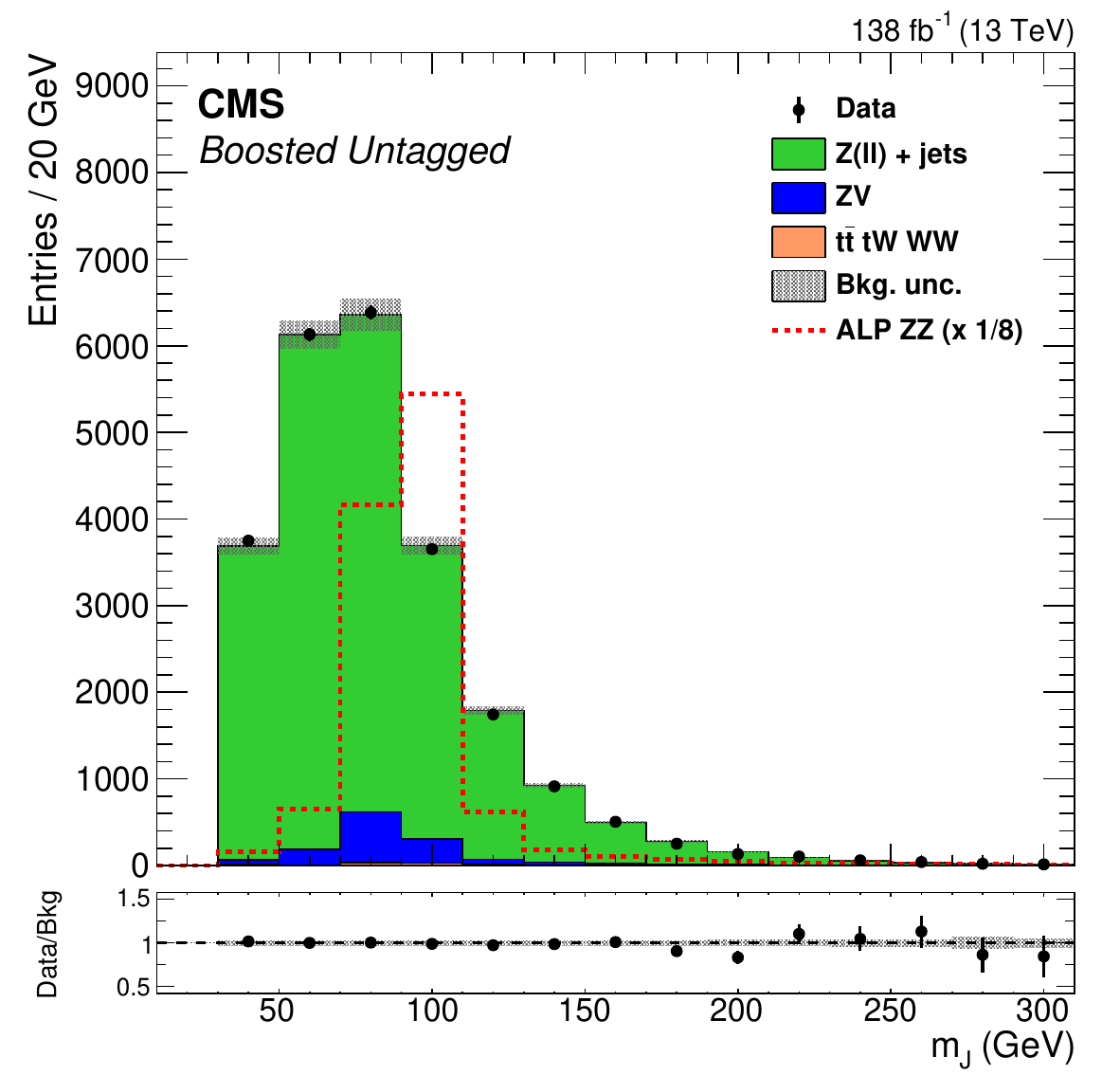}
\includegraphics[width=0.49\textwidth]{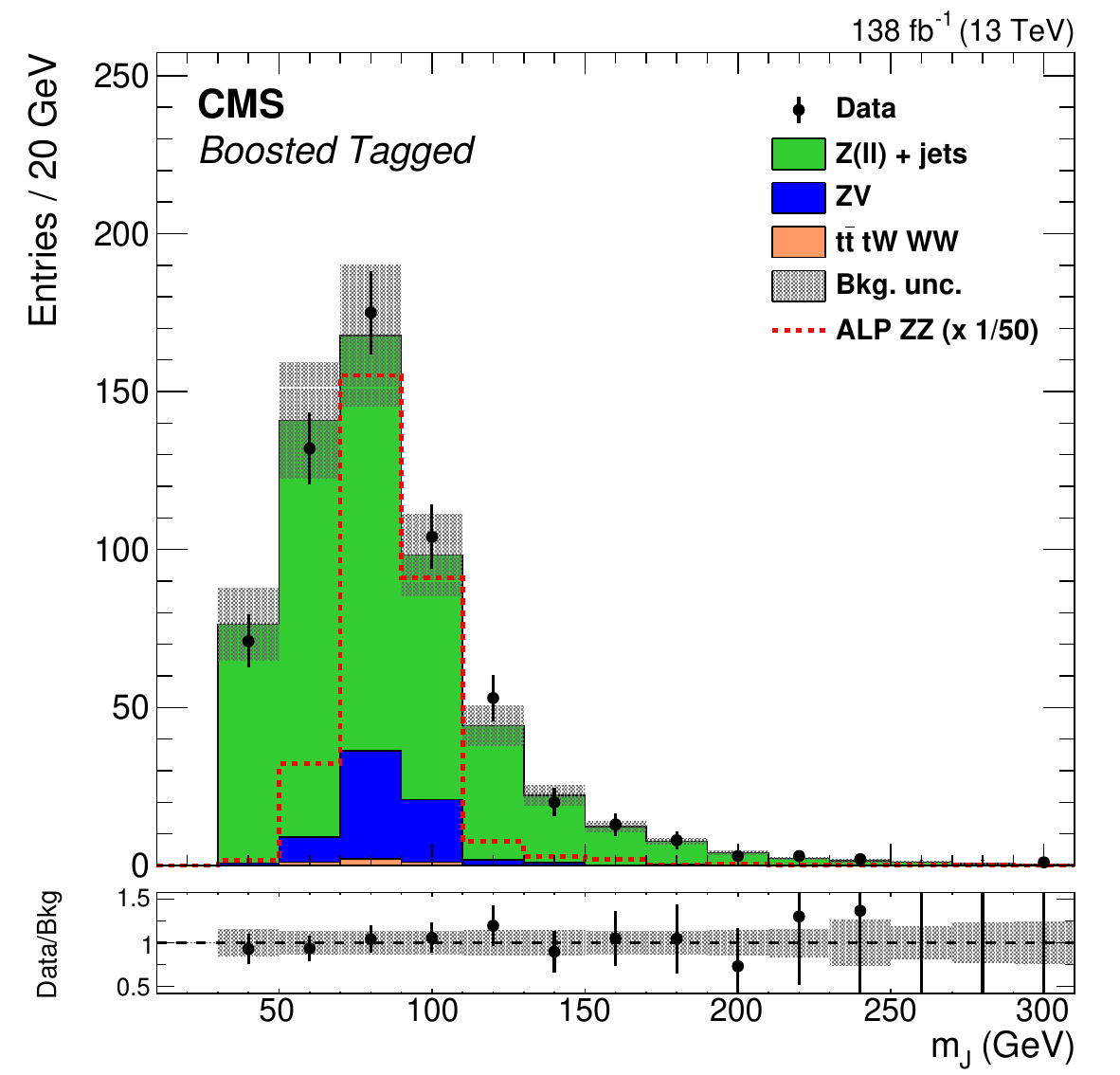}
\includegraphics[width=0.49\textwidth]{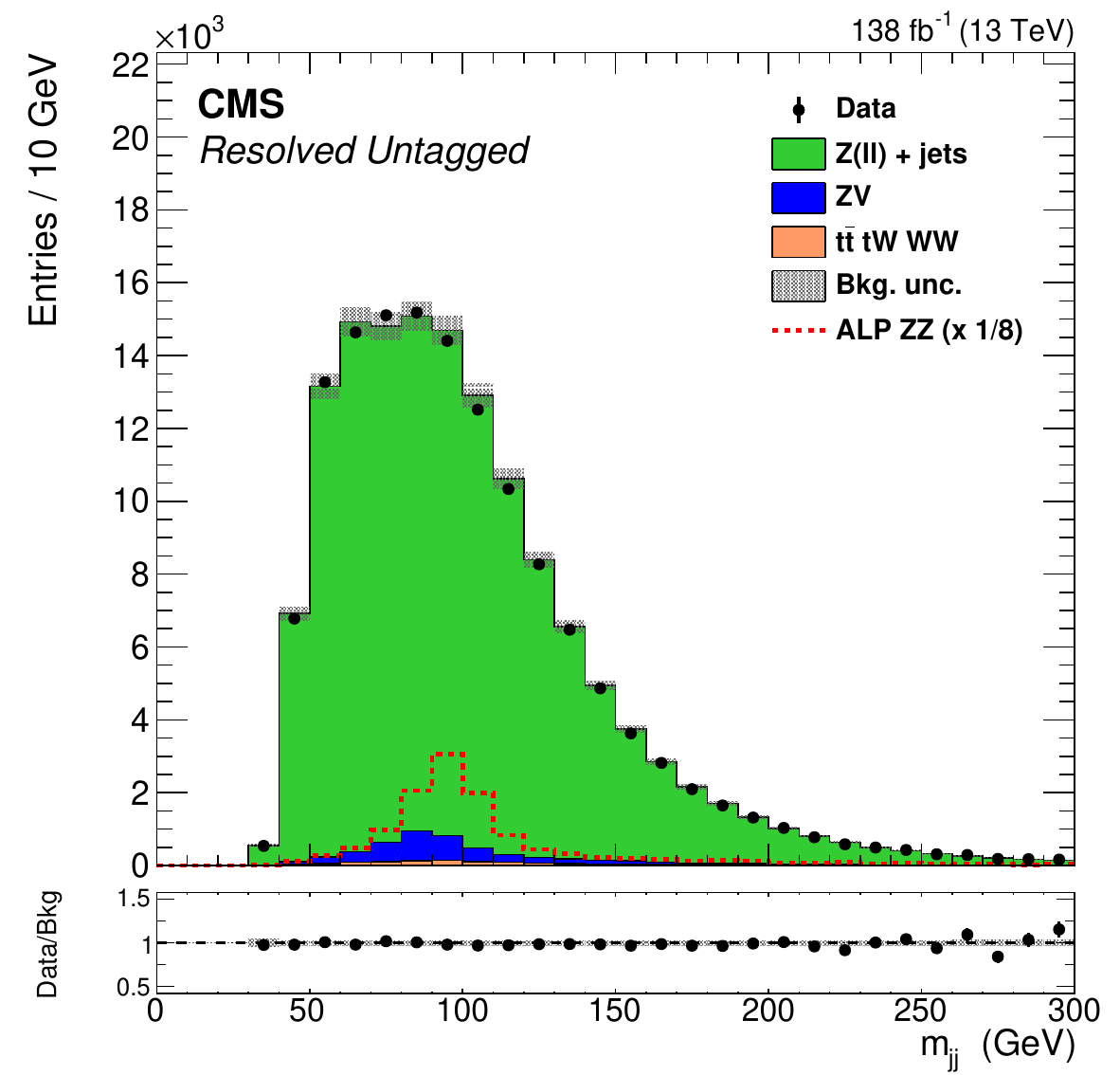}
\includegraphics[width=0.49\textwidth]{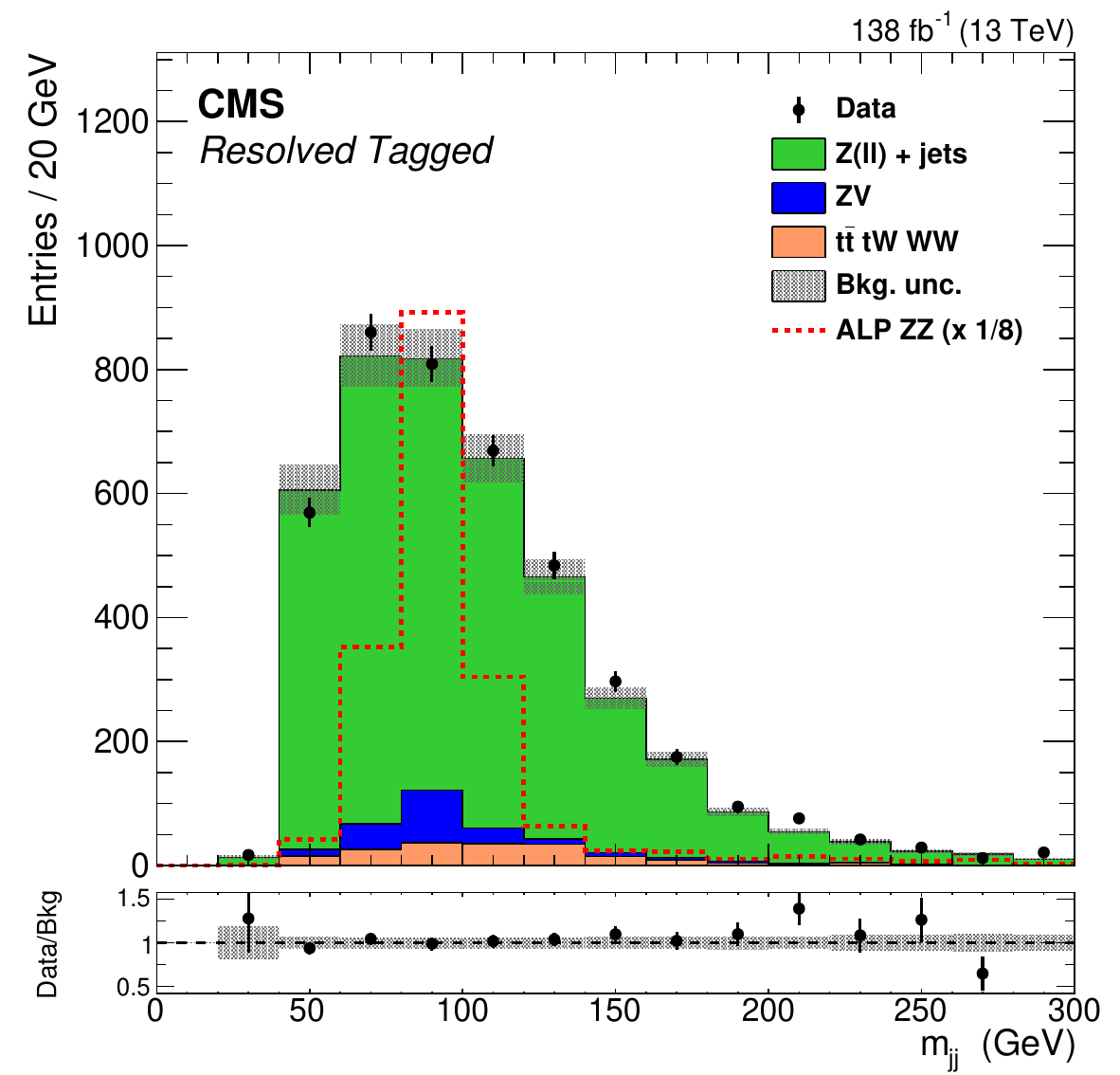}
\caption{
Distributions of the merged jet \mj (upper) and the dijet \mjj (lower) for the untagged (left) and tagged (right) categories.
The distributions include events in the signal regions SR1 and SR2 and in the sideband SB; the corresponding boundaries have been defined in the text. 
The gray band shows the statistical and systematic uncertainties in the background. 
Background normalizations are derived from the final fit to the data.
The red dashed histograms correspond to a hypothetical linear ALP with $1\TeV^{-1}$ couplings to gluons and \ZZ, and $\falp = 3\TeV$; the cross sections have been multiplied by the factors indicated in the legends for better visibility. The lower panels show the ratio of data to background.
}
\label{fig:vmass}
\end{figure*}

In total, eight categories are defined by the combination of electrons/muons, boosted/resolved, and tagged/untagged.
Table~\ref{table:SelCtg} summarizes the selection and categorization requirements of the analysis.
Figure~\ref{fig:vmass} shows the merged jet \mj and dijet \mjj distributions for the untagged and tagged categories in the signal and sideband regions.

\begin{table}[htbp]
\topcaption{Summary of selection requirements and categorization.}\label{table:SelCtg}
\renewcommand{\arraystretch}{1.2}
\centering
\begin{tabular}{lcc}
\multirow{2}{*}{Observable}  	& \multicolumn{2}{c}{Event category}  \\[-3pt]
				& Resolved & Boosted \\ 
\hline
Leptons			& \multicolumn{2}{c}{Same flavor, opposite charge} \\
$\pt(\Pell)$ (\GeVns)  	& $>$40  	& $>$40 \\ 
$\abs{\eta(\Pell)}$	& $<$2.4 	& $<$2.4 \\ 
\mll (\GeVns)		& [76, 106] 	& [76, 106] \\ 
\abs{\eta(\Pj)} or \abs{\eta(\PJ)}  	& $<$2.4  	&  $<$2.4 \\ 
$\DR(\Pell\PJ)$ (AK8)   	& \NA  		& $>$0.8 \\
$\DR(\Pell\Pj)$ (AK4)   	& $>$0.4 	& \NA \\
$\DR(\Pj\Pj)$		& $<$1.5	& \NA \\
$\pt(\Pell\Pell)$ (\GeVns)  & $>$150       & $>$200 \\
$\pt(\Pj\Pj)$ or $\pt(\PJ)$ (\GeVns)    & $>$150 	& $>$200 \\
PUPPI $\tauto$             & \NA	& $<$0.4 (2016), $<$0.45 (2017, 2018)\\
SR1 \mjj or \mj (\GeVns)	& [65, 110] 	& [65, 105] \\ 
SR2 \mjj or \mj (\GeVns)	& [95, 135]	& [95, 135] \\ 
SB \mjj or \mj (\GeVns)	& [30, 65] $\bigcup$ [135, 180] & [30,65] $\bigcup$ [135,300]  \\ 
\PQb tagging		& Untagged / Tagged [L+M]	& Untagged / Tagged [L+M] \\
\end{tabular}
\end{table} 

\newpage
\section{Background estimation}\label{sec:bkg}

Most events in the selected \ZV and \ZH candidate samples come from \Zjets production. A genuine \Z boson
decaying into $\Pell\Pell$, along with a high-\pt merged jet or dijet pair, may easily yield a signal-like combination,
even though the jets do not originate from a boson decay.
The features of the \Zjets background are studied using simulated samples.
The limited number of simulated events remaining in the \PQb-tagged categories results in sizable statistical fluctuations in the predicted distributions of the diboson mass \mZV and \mZH for this background.
It has been observed, however, that within simulation uncertainties, the shape of the \Zjets mass distribution is the same for events with and without \PQb-tagged jets.
Therefore, the \Zjets shapes in the \PQb-tagged categories are described using the \mZV and \mZH shapes obtained from the simulation without making any \PQb tag requirements.

\begin{figure}[htbp]
\centering
\includegraphics[width=0.49\textwidth]{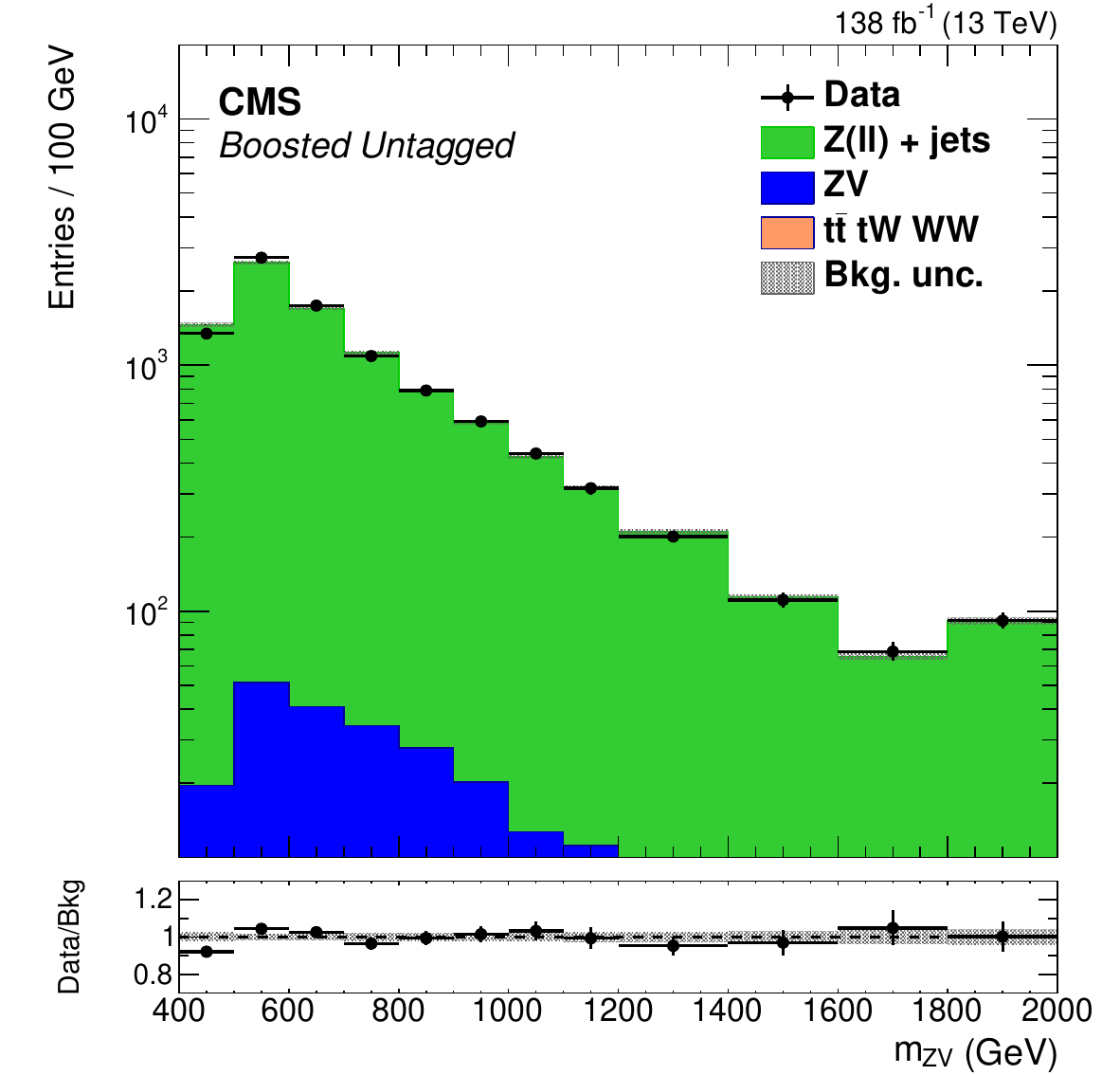}
\includegraphics[width=0.49\textwidth]{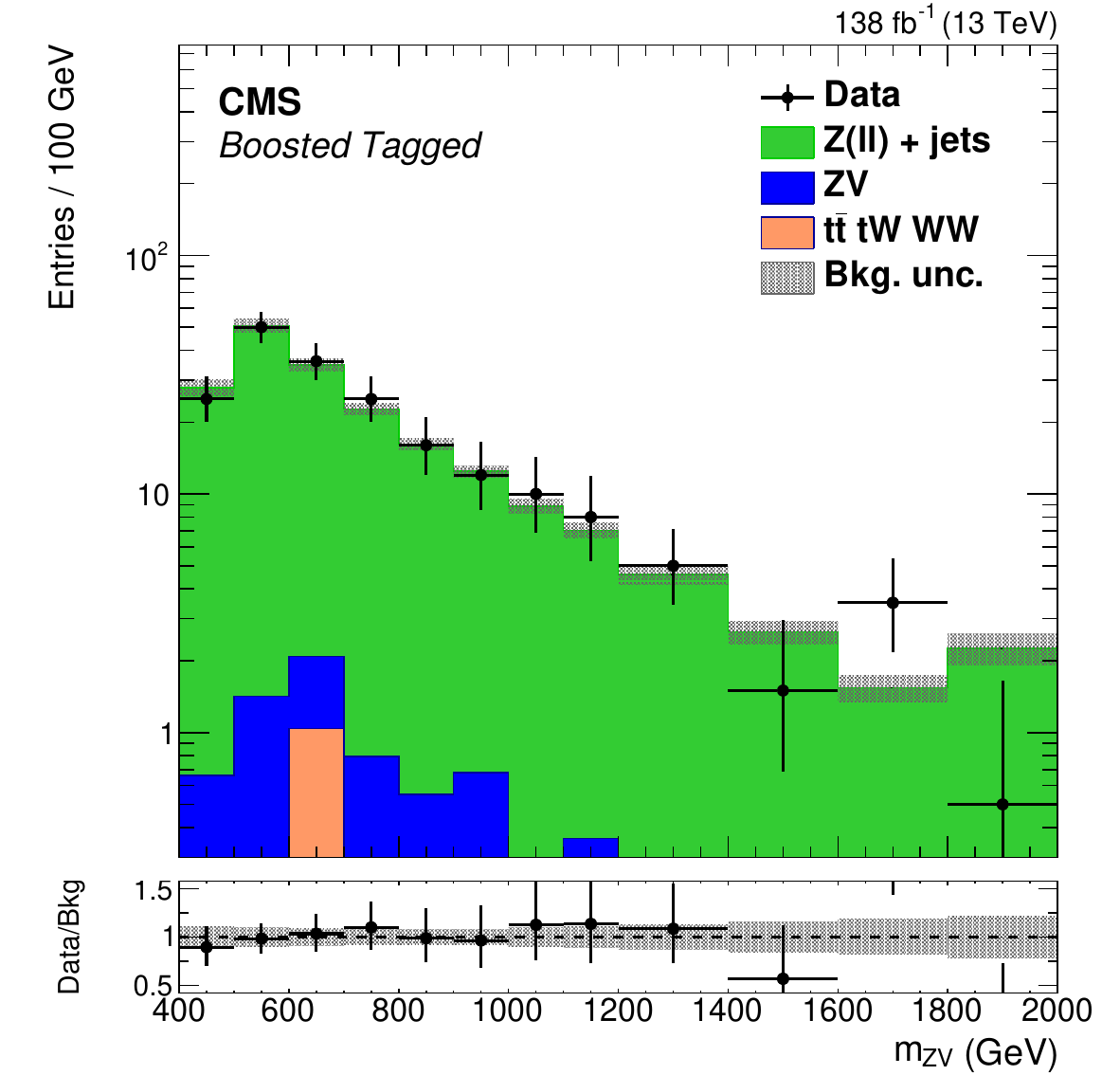} \\
\includegraphics[width=0.49\textwidth]{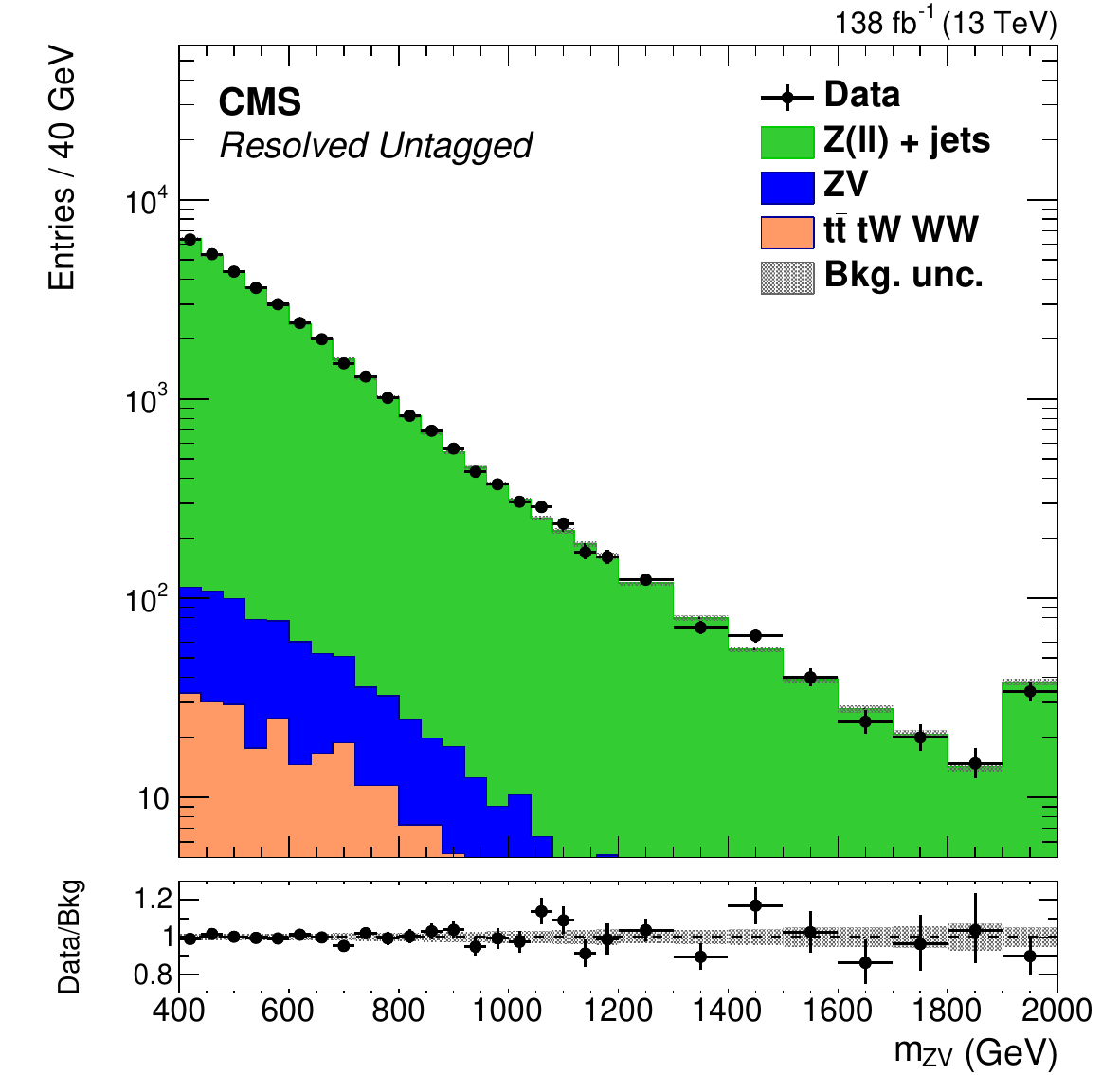}
\includegraphics[width=0.49\textwidth]{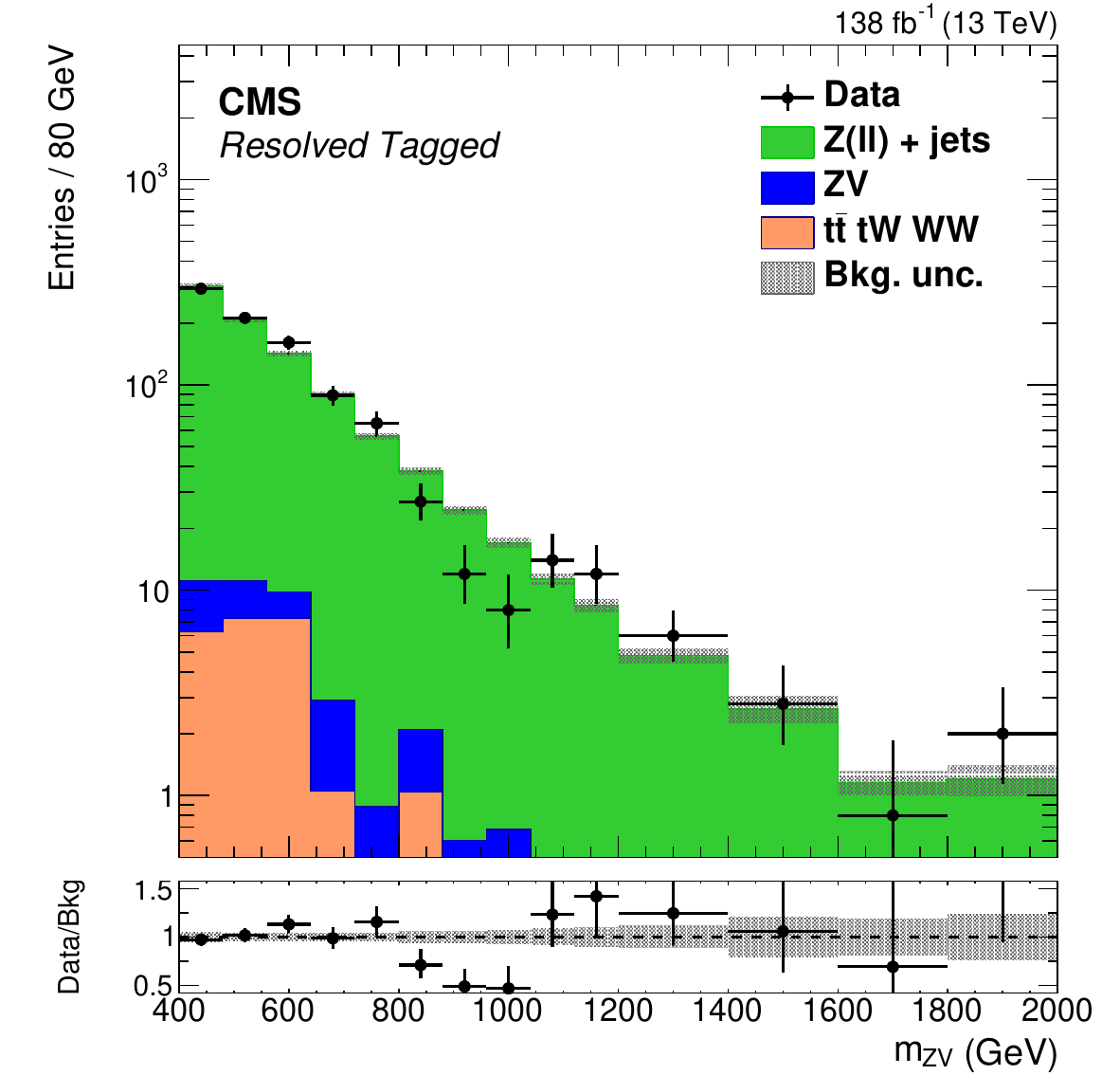} \\
\caption{The sideband diboson mass distributions for the boosted \V (upper), resolved \V (lower), untagged
(left), and tagged (right) categories after fitting the sideband data alone. 
The last bin includes events with \mZV values up to 3000\GeV.
The points show the data, while the filled histograms show the background contributions.
The gray band indicates the statistical and post-fit systematic uncertainties in the
normalization and shape of the background. The lower panel shows the ratio of data to background.
}
\label{fig:XMass_SBfit}
\end{figure}

The modeling of the \Zjets background is obtained from simulation and corrected using data.
In the final fit to the data, the \Zjets background normalization in the signal region, SR1 or SR2, is constrained by the observed yield in the SB.
The shape of the \Zjets mass distribution is determined from the simulation, with a correction applied to improve the agreement with the data in the SB region.
A linear function depending on one single parameter $s$,
$\corr(\mZX, s) = 1 + s (\mZX - 500\GeV) / (500\GeV)$,
is sufficient to make the invariant mass distributions agree within their uncertainties.
This procedure is applied independently to each boosted/resolved and untagged/tagged category.
Allowing for different linear corrections in the untagged and tagged categories takes into account possible differences in the shape of the \Zjets background.
Figure~\ref{fig:XMass_SBfit} shows fits to the SB diboson mass distributions including the linear correction with $s$ at its fitted value. Fitted values of $s$ are in the range of $-0.05$ to $+0.10$, with $s=0$ corresponding to no correction.
The relative uncertainties in the SB-fitted slope parameter $s$ are in the range 30--100\%, depending on the category.
In the final fit of each category, the slope parameter $s$ is allowed to float, constrained by the data in the signal and sideband regions simultaneously.
The corrections to the \Zjets shapes derived from the SB and the SR are found to be mutually consistent.
Statistical uncertainties associated with the simulated \Zjets distributions are also taken into account in the fit. The fits in the boosted \V categories include the peaking region of the background; 
Fig.~\ref{fig:XMass_SBfit} shows that the SB data in this particular region are described well by the fit.
The systematic uncertainty in the estimation of the \Zjets background is given by the uncertainties in the slope correction parameter $s$ and in the normalization.

Dilepton backgrounds that do not contain a leptonic \PZ boson decay are estimated from data using \EM events passing the analysis selection.
This approach accounts for top quark pair production (\ttbar), $\WW + \text{jets}$, $\Ztott + \text{jets}$, single top quark, and hadrons misidentified as leptons; we collectively refer to this set of backgrounds as \tplusX.
The relative normalization and shape of the diboson mass distributions for \ee and $\PGm\PGm$ events with respect to \EM events have been estimated from data on a top quark-enriched control sample and shown to be consistent with lepton flavor symmetry expectations.
The  \tplusX background  contamination represents 0.8 (0.9)\% and 3 (4)\% of the total background in the untagged and tagged categories of the resolved SR1 (SR2) selection, while the boosted SR1 (SR2) selection has a contamination of 0.5 (0.7)\% and 0.8 (1)\% in the untagged and tagged categories.

The diboson background (\ZZ and \ZW, with \Ztoll) is estimated directly from simulation.
The diboson contamination represents 4 (3)\% and 8 (4)\% of the total background in the untagged and tagged categories of the resolved SR1 (SR2) selection, while the boosted SR1 (SR2) selection has a contamination of about 8 (4)\% and 20 (11)\% in the untagged and tagged categories. The contribution to the background of SM \ZH production, with \Ztoll, has been found to be negligible.

\begin{figure*}[htbp]
\centering
\includegraphics[width=0.49\textwidth]{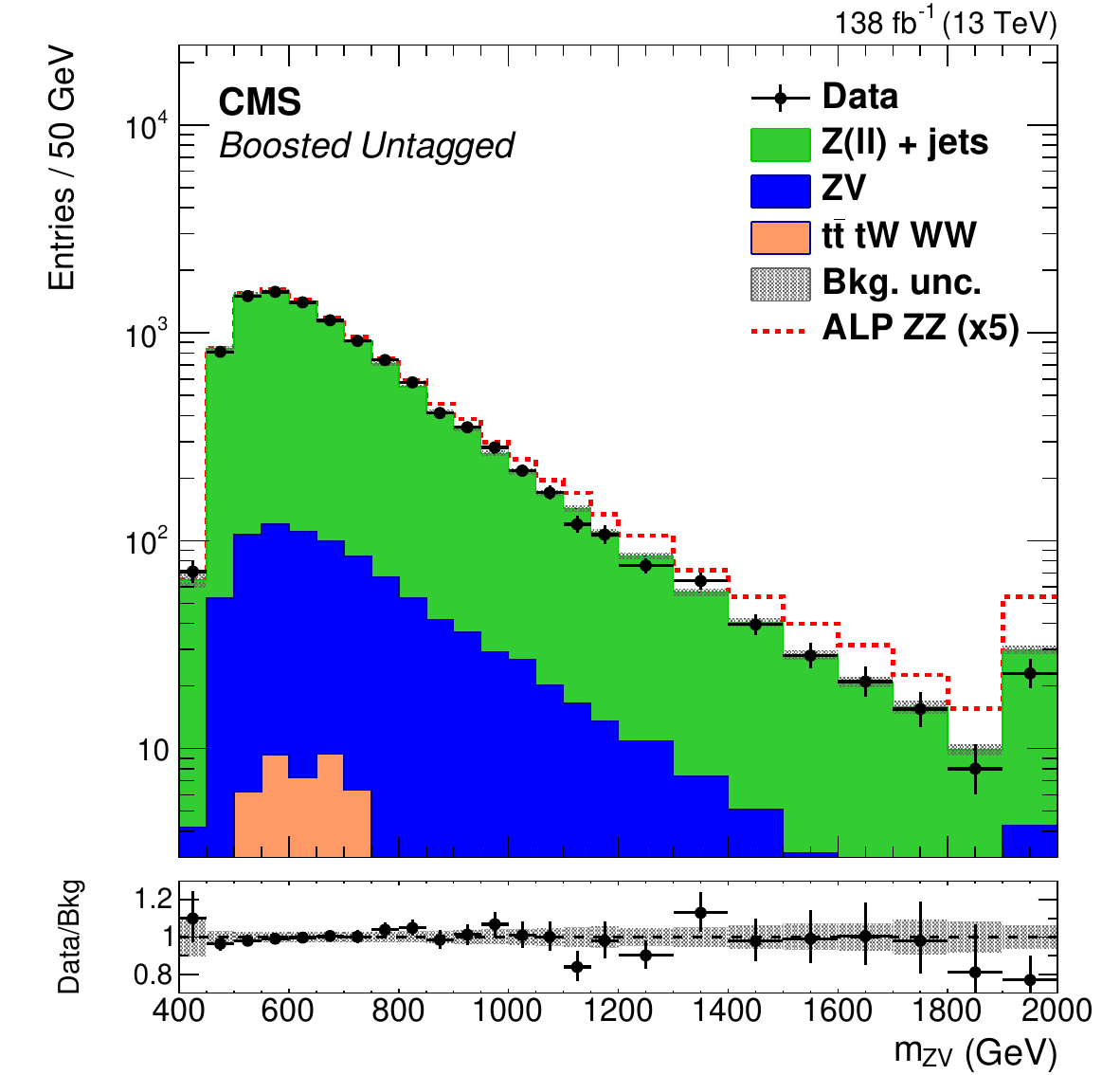}
\includegraphics[width=0.49\textwidth]{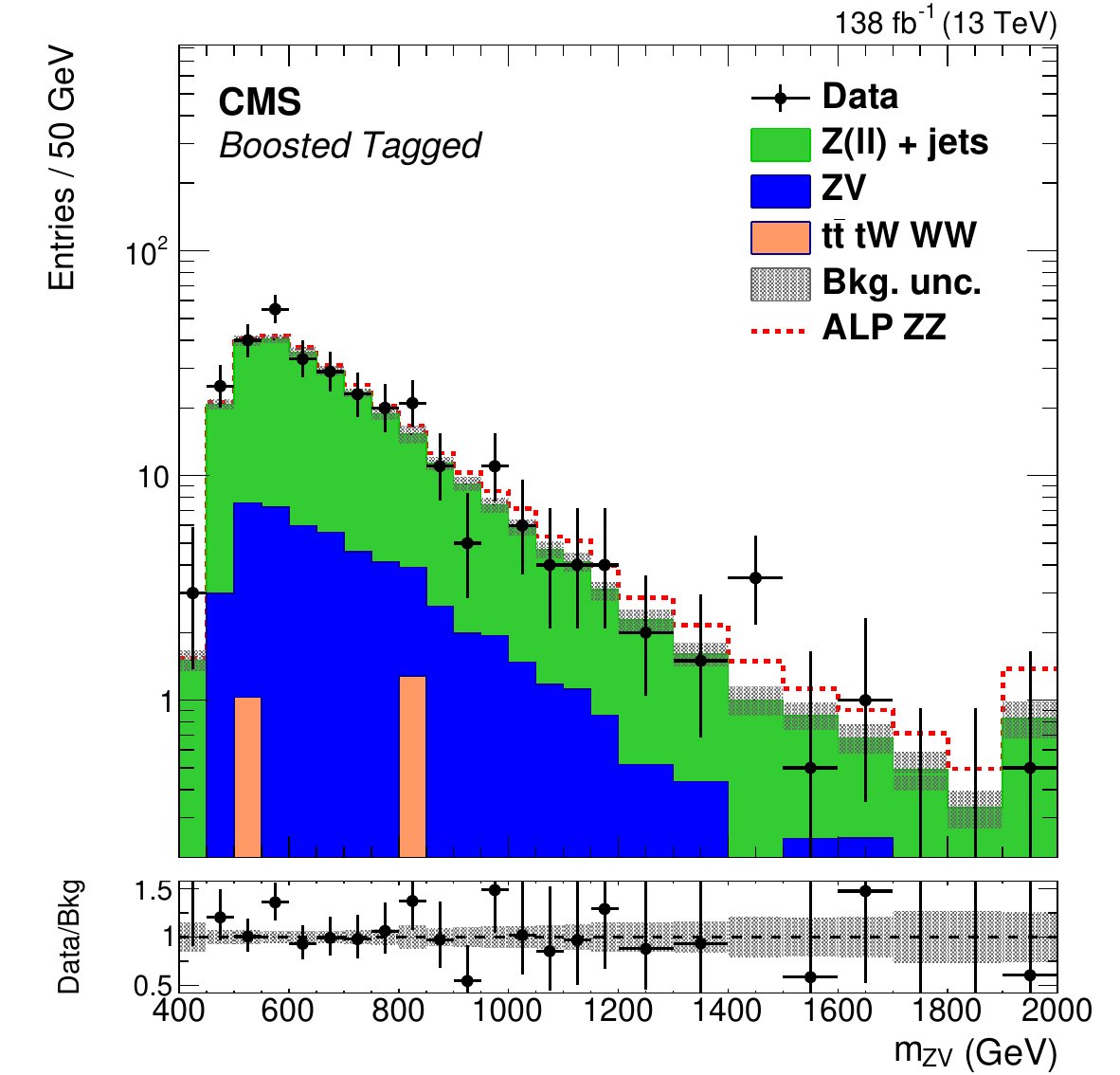}
\includegraphics[width=0.49\textwidth]{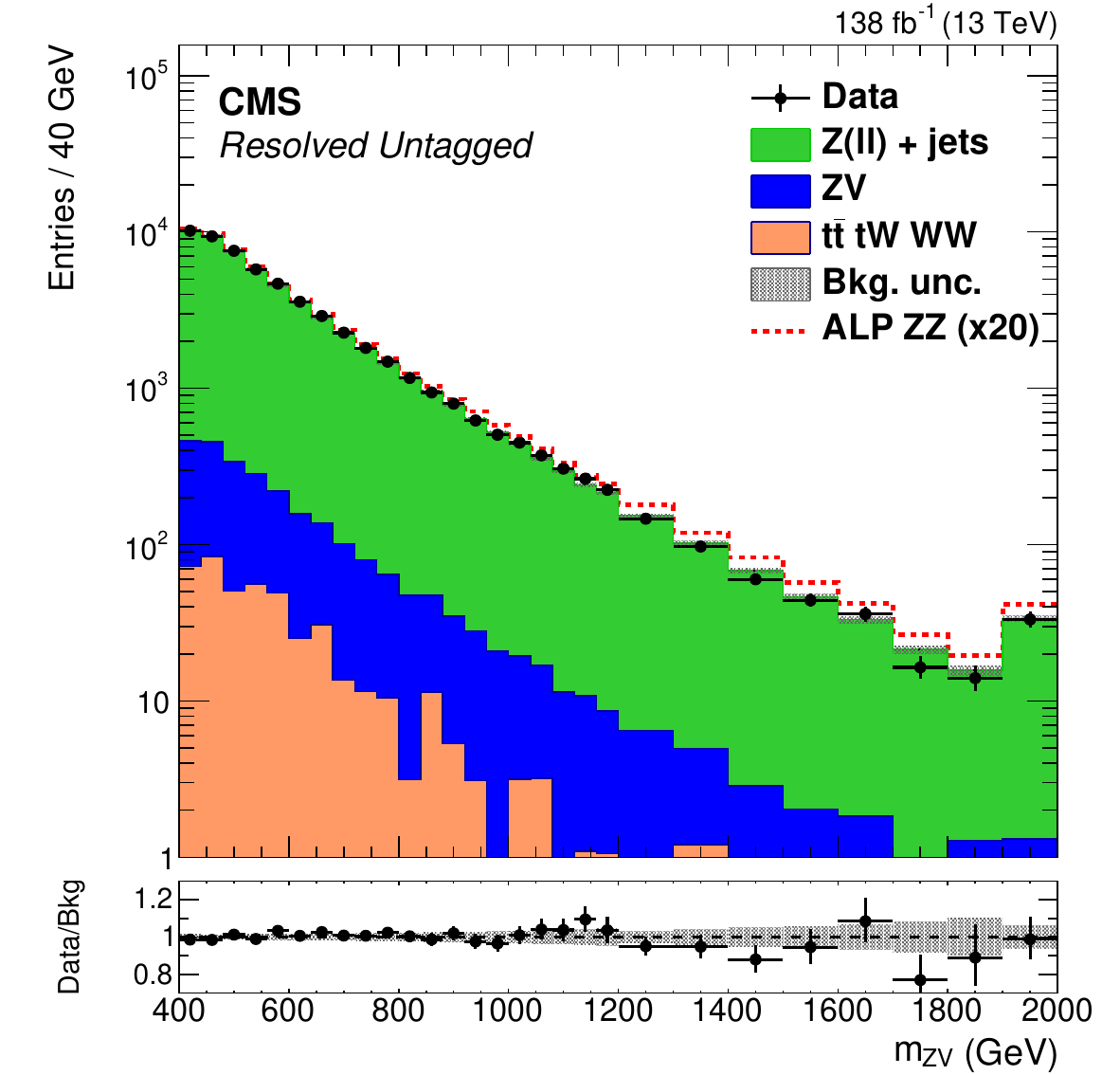}
\includegraphics[width=0.49\textwidth]{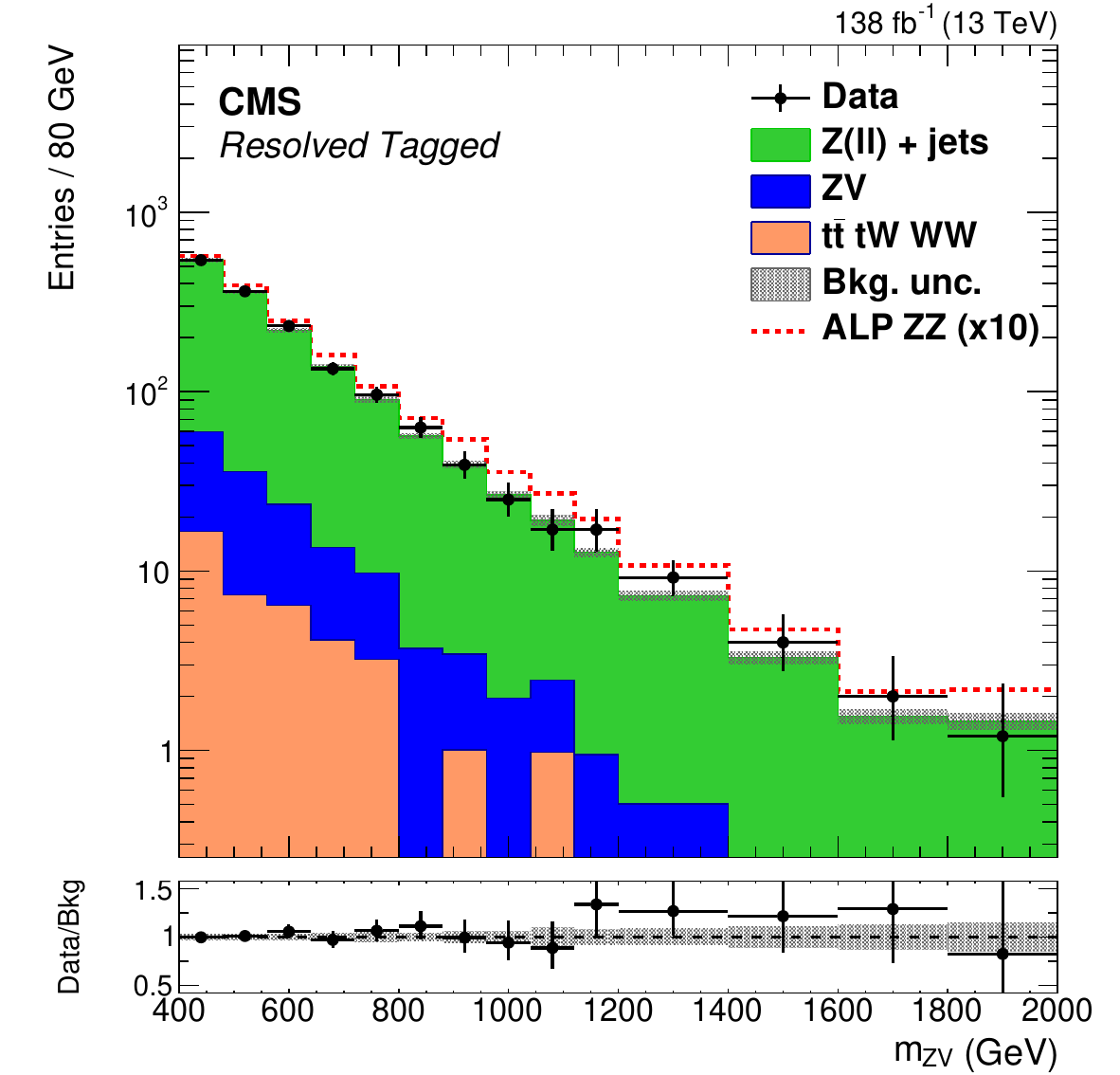}
\caption{The SR1 \mZV distributions for the boosted \V (upper), resolved \V (lower), untagged
(left), and tagged (right) categories
after fitting the signal and sideband regions
with a model comprising signal (ALP linear \ZZ) plus background.
The last bin includes events with \mZV values up to 3000\GeV.
The points show the data, while the filled histograms show the
background contributions. The signal is represented by the red dashed histogram, normalized to the observed 95\% confidence level
cross section limit at $\falp = 3\TeV$; the cross sections have been multiplied by the factors indicated in the legends for better visibility.
The gray band indicates the statistical and post-fit systematic uncertainties in the normalization and shape of
the background. The lower panels show the ratio of data to background.
}
\label{fig:ALPl_SR1fit}
\end{figure*}

\begin{figure*}[htbp]
\centering
\includegraphics[width=0.49\textwidth]{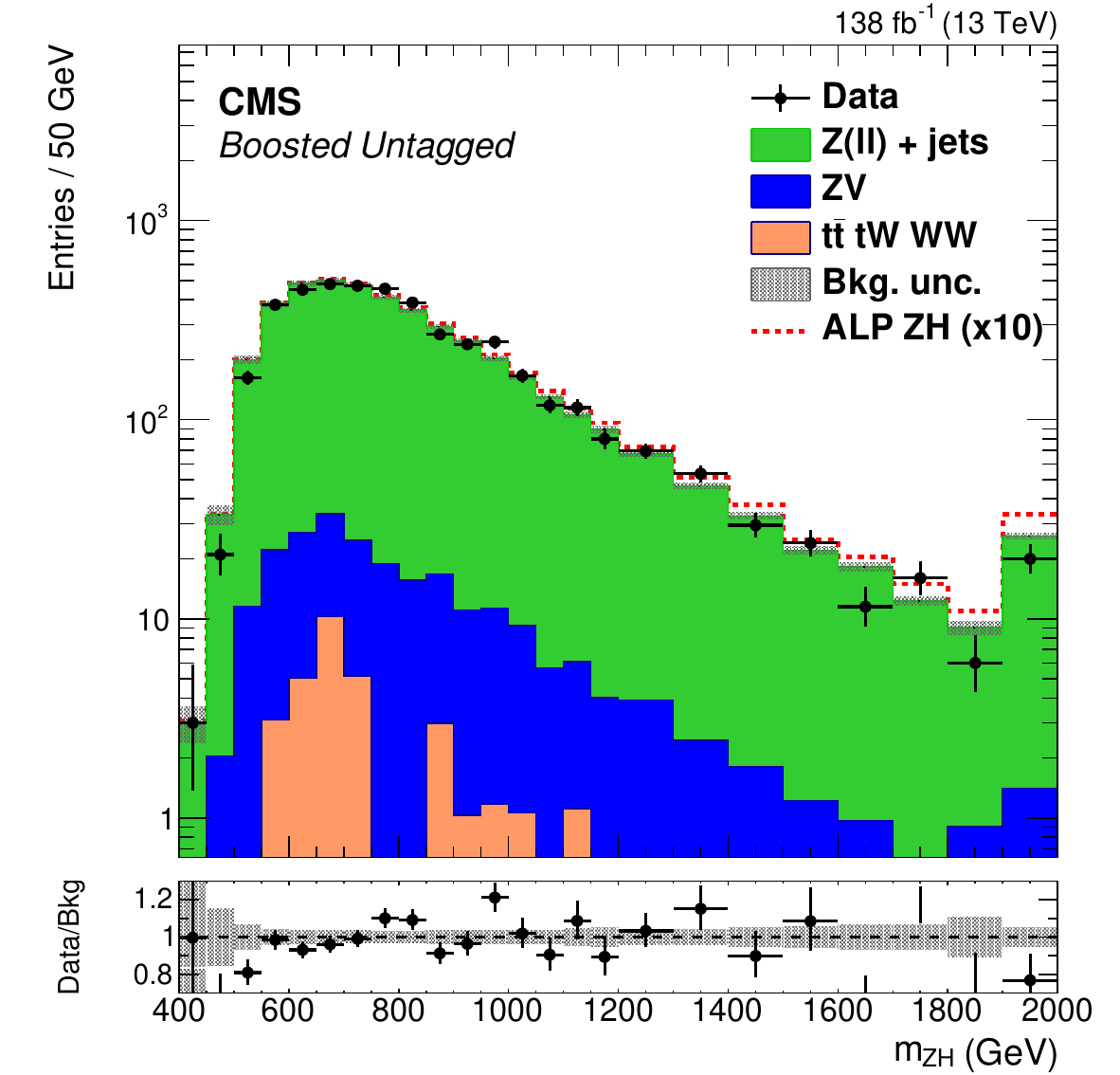}
\includegraphics[width=0.49\textwidth]{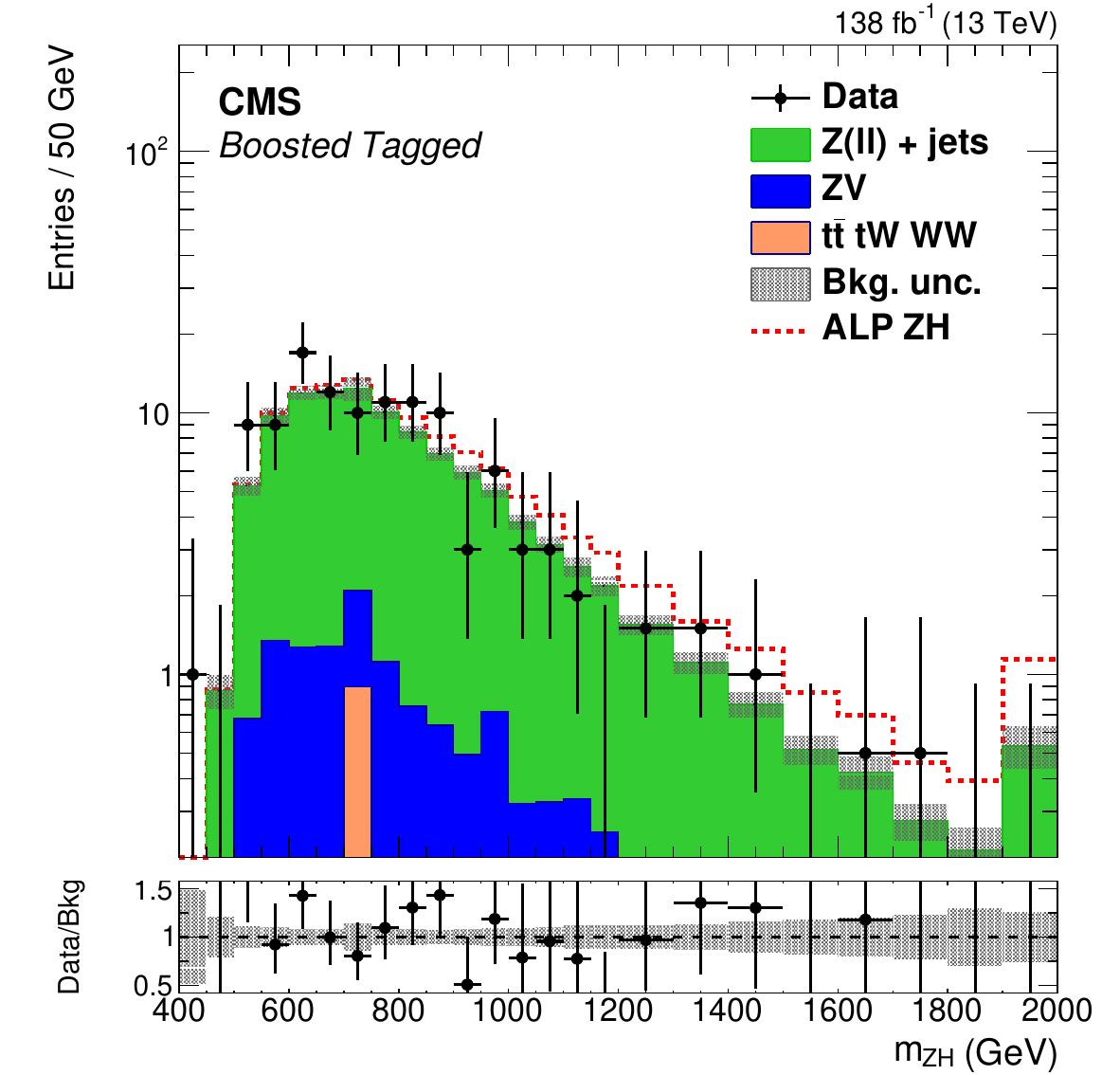}
\includegraphics[width=0.49\textwidth]{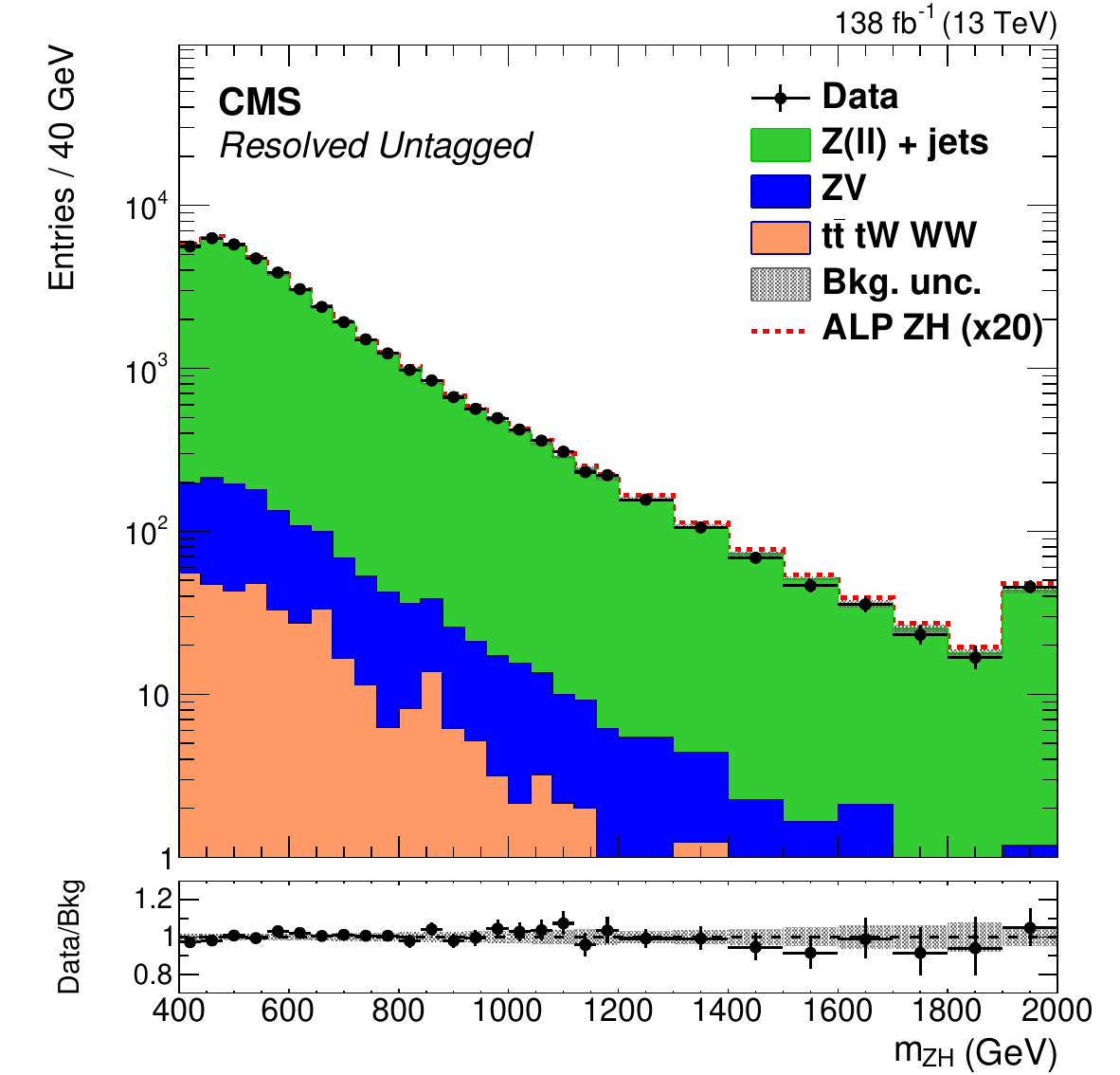}
\includegraphics[width=0.49\textwidth]{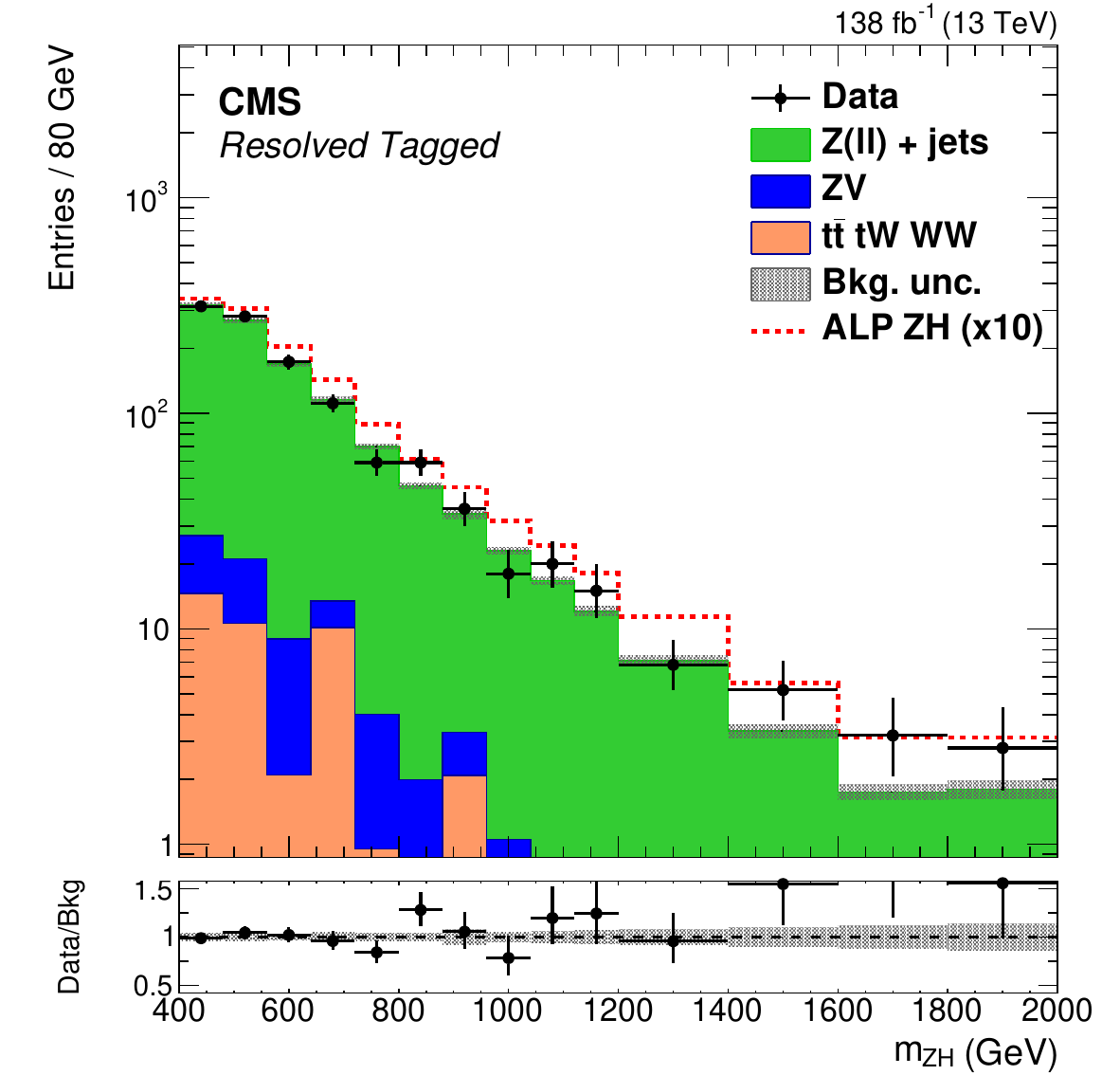}
\caption{The SR2 \mZH distributions for the boosted \PH (upper), resolved \PH (lower), untagged
(left), and tagged (right) categories
after fitting the signal and sideband regions
with a model comprising signal (ALP chiral \ZH) plus background.
The last bin includes events with \mZH values up to 3000\GeV.
The points show the data, while the filled histograms show the
background contributions. The signal is represented by the red dashed histogram, normalized to the observed 95\% confidence level
cross section limit at $\falp = 3\TeV$; the cross sections have been multiplied by the factors indicated in the legends for better visibility.
The gray band indicates the statistical and post-fit systematic uncertainties in the normalization and shape of
the background. The lower panels show the ratio of data to background.
}
\label{fig:ALPc_SR2fit}
\end{figure*}

The SR1 \mZV and SR2 \mZH distributions for the boosted and resolved categories are shown in Figs.~\ref{fig:ALPl_SR1fit} and~\ref{fig:ALPc_SR2fit}, respectively.

\section{Systematic uncertainties}\label{sec:sys}

The systematic uncertainties influence both the normalization and shape of the background and signal distributions in the analysis.

The uncertainty in the background shape, which is the dominant systematic effect, is evaluated by comparing data and simulation after the fit in the signal and SB regions, and the residual shape difference is treated as an uncertainty.
The background shape correction procedure has been explained in Section~\ref{sec:bkg}.
The impacts of these uncertainties are quantified by calculating the change on the fitted signal cross section when a given parameter is displaced by ${\pm}1$ standard deviation from its post-fit value, divided by the total uncertainty in the fitted signal cross section. 
The impacts of the background shape uncertainty in the boosted untagged, boosted tagged, resolved untagged, and resolved tagged categories are 11, 13, 3, and 3\%, respectively, for a bulk graviton with a mass of 1\TeV. 
For an ALP linear \ZZ (ALP chiral \ZH) signal, the corresponding impacts are 42 (9), 42 (44), 16 (7), and 16 (23)\%, respectively.

Uncertainties associated with the description in simulation of the trigger efficiencies, as well as the uncertainties in the efficiency for lepton reconstruction, identification, and isolation, are extracted from dedicated studies of events with leptonic \PZ boson decays, and amount to 1.5\% for muons and 2\% for electrons.
The uncertainties in the lepton momentum and energy scales are propagated to the signal shape and normalization; the largest impact on the normalization is 0.2\%.

Uncertainties in the jet energy scale and resolution~\cite{CMS-PAS-JME-16-003} affect both the normalization and the shape of the signal samples.
The momenta of the reconstructed jets are varied according to the uncertainties in the jet energy scale and resolution, and the selection efficiencies and signal shapes are reevaluated using these modified samples. This procedure results in a change of 1--2\% in the normalization of the resolved categories; the largest change in the normalization of the boosted categories is 0.3\%.
These sources of uncertainty are responsible for the largest signal shape systematic uncertainty.

The uncertainty in the boosted \V boson identification efficiency and the groomed mass scale and resolution~\cite{Khachatryan2014} are responsible for a 7\% signal normalization uncertainty in the merged category of the analysis.
These are measured in data and simulation in an almost pure selection of semileptonic \ttbar events where boosted \PW bosons produced in top quark decays are separated from the combinatorial \ttbar background by means of a simultaneous fit to the soft-drop merged-jet mass.
The uncertainties in the soft-drop mass scale and resolution are propagated to the groomed jet mass, and the impact on the selection efficiency of signal and \ZV background is taken into account.
The simulated \tauto distributions have been found to be similar for boosted \PW, \PZ, and \PH bosons, allowing the same signal normalization uncertainty to be applied to all boosted bosons.
An additional uncertainty affecting the signal normalization is included to account for the extrapolation of the uncertainties extracted from a \ttbar sample at typical jet \pt of 200\GeV to higher regimes,
yielding an uncertainty of 2.6--6\%.
Extrapolation uncertainties are estimated by comparing the results from simulated event samples using \HERWIG~\cite{Bahr:2008pv} and \PYTHIA. 

The scale factors (SFs) for the \PQb tagging efficiencies and their corresponding systematic uncertainties~\cite{Sirunyan:2018btv} are measured in data, as functions of \pt, using samples enriched in \PQb quark content, and their propagation to the signal region of the analysis produces an uncertainty of 12\% in the tagged boosted category and 4\% in the tagged resolved category. The uncertainties in the mistag efficiency SFs are also considered; the uncertainties in the \PQb tagging and mistag efficiencies are treated as anticorrelated between the tagged and untagged categories.

For the \tplusX background, a 4\% uncertainty is estimated by comparing the yield of \EM events with $\ee + \PGm\PGm$ events in a top quark-enriched control region.

The different shapes of the \mj (\mjj) distributions in the NLO and LO \Zjets simulations induce a 3 (5)\% change in the ratio of the SR to SB normalizations in the boosted (resolved) categories. This difference is taken as a systematic uncertainty.

The effect of varying the factorization and renormalization scales by a factor of 2 is propagated to the signal shape and normalization.
The impact on the signal acceptance is evaluated to be 0.1--0.3\%, depending on the signal mass and analysis category.
Varying the factorization and renormalization scales in the \Zjets simulation changes the shape of the distribution in the same way as the linear correction, by an amount that is less than the correction itself. The effect is therefore absorbed by the fit, and an explicit associated uncertainty is not required.

The systematic uncertainty associated with the PDFs used to generate the simulated samples is evaluated by varying the NNPDF~3.0/3.1 PDF set within its uncertainties, following the prescription in Ref.~\cite{Butterworth:2015oua}, and its effect is propagated to the signal shape and normalization, resulting in a measured uncertainty of 0.3\% in the resolved categories and 1.5\% in the boosted categories.

The effect of varying the scales and PDFs on the shape of the ALP distribution is also evaluated, but as it is found to be small (at most 3\% in the tail of the distribution), it is neglected in the final fit to the data.

Additional systematic uncertainties affecting the normalization of backgrounds and signal from the contributions of pileup events and the integrated luminosity~\cite{CMS-LUM-17-003,CMS-PAS-LUM-17-004,CMS-PAS-LUM-18-002} are also considered.

The complete list of background and signal normalization systematic uncertainties in the analysis is reported in Table~\ref{table-syst}.

\begin{table}[!htb]
\centering
\topcaption{
Summary of systematic uncertainties, quoted in percent, affecting the normalization of background and signal samples. Where a systematic uncertainty depends on the signal \ZV or \ZH channel or mass, the smallest and largest values are reported. In the case of a systematic uncertainty applying only to a specific background source, the source is indicated in parentheses. Systematic uncertainties too small to be considered are written as ``$<$0.1'', while a dash (\NA) represents uncertainties not applicable in the specific analysis category.
\label{table-syst}
}
\begin{tabular}{lcccc}
                       & \multicolumn{2}{c}{Boosted}   & \multicolumn{2}{c}{Resolved} \\
Source                 & Background   & Signal         & Background & Signal     \\
\hline
Integrated luminosity  & \multicolumn{2}{c}{1.8}       & \multicolumn{2}{c}{1.8} \\
Electron trigger and ident.& \multicolumn{2}{c}{2.0}       & \multicolumn{2}{c}{2.0} \\
Muon trigger and ident.    & \multicolumn{2}{c}{1.5}       & \multicolumn{2}{c}{1.5} \\
Electron energy scale  & 0.8          & $<$0.1--0.2    & 0.9        & $<$0.1     \\
Muon momentum scale    & 0.5          & $<$0.1--0.1    & 0.6        & $<$0.1     \\
Jet energy scale       & 1.0          & $<$0.1--0.1    & 2.8        & 0.1--1.9   \\
Jet energy resolution  & 0.3          & $<$0.1--0.3    & 0.3        & 1.0        \\
\V/\PH identification (\tauto)  & 5 (\ZV)      & 5              & \NA        & \NA        \\
\V/\PH identification (extrap.) & \NA          & 2.6--6.0       & \NA        & \NA        \\
\V/\PH mass scale      & 0.6 (\ZV)    & 0.4--0.8       & \NA        & \NA        \\
\V/\PH mass resolution & 5.0 (\ZV)    & 5.0--6.0       & \NA        & \NA        \\
\PQb tag SF, untagged  & 0.1          & 1.0--7.4       & 0.1        & 0.7--2.2   \\
\PQb tag SF, tagged    & 12           & 12             & 3.6        & 4          \\
Mistag SF, untagged    & 0.3          & $<$0.1--0.2    & 0.2        & 0.1        \\
Mistag SF, tagged      & 3.5          & 0.1--0.3       & 3.8        & 0.4--1.0   \\
SM \ZV production      & 12           & \NA            & 12         & \NA        \\
\tplusX normalization  & 4 (\EM)      & \NA            & 4 (\EM)    & \NA        \\
SR-to-SB norm. ratio   & 3 (\Zjets)   & \NA            & 5 (\Zjets) & \NA        \\
PDFs                   & \NA          &  1.5--1.6      & \NA        & 0.3--1.1   \\
Renorm. and fact. scales & \NA        &  0.1--0.3      & \NA        & 0.2--0.3   \\
Pileup                 & 0.5          &  0.1--0.2      & 0.1        & 0.1--0.2   \\
MC statistics, untagged& 0.3          &  0.7           & 0.2        & 1          \\ 
MC statistics, tagged  & 2            &  1.5           & 1.5        & 2          \\ [2ex]
Total, untagged        & 4            & 8--13          & 6          & 3--4       \\
Total, tagged          & 13           & 14--16         & 8          & 5--6       \\          
\end{tabular}
\end{table}

\section{Results and interpretation}\label{sec:res}

The signal selection efficiencies obtained from simulation are presented in Table~\ref{table:sigeff}. The efficiencies are relative to simulated events in which one \PZ boson decays to an electron or muon pair, the \PW boson or second \PZ boson decays to quarks, and the \PH boson decays according to the branching fractions of the SM.

\begin{table*}[htb!]
\centering
\topcaption{Selection efficiencies in percent for the bulk graviton, \PWpr, and ALP linear and chiral models. The efficiencies are relative to simulated events in which one \PZ boson decays to an electron or muon pair, the \PW boson or second \PZ boson decays to quarks, and the \PH boson decays according to the branching fractions of the SM.}
\label{table:sigeff}
\cmsTable{
\begin{tabular}{lcccccccc}
Analysis &  500\GeV &  600\GeV &  800\GeV &  1000\GeV &  1200\GeV  &  1400\GeV & 1600\GeV & 1800\GeV\\   
\hline
\multicolumn{9}{l}{Bulk graviton}\\ 
Boosted  &  8 & 22 & 33 & 37 & 38 & 38 & 37 & 36 \\
Resolved & 21 & 15 & 12 &  9 &  6 &  4 &  3 &  2 \\
Combined & 29 & 37 & 44 & 46 & 44 & 42 & 40 & 38 \\ [\cmsTabSkip]

\multicolumn{9}{l}{\PWpr}\\
Boosted  & 12 & 25 & 35 & 39 & 40 & 39 & 39 & 37 \\
Resolved & 14 & 10 &  9 &  6 &  4 &  3 &  2 &  1 \\
Combined & 26 & 35 & 43 & 45 & 44 & 42 & 40 & 38 \\ [\cmsTabSkip]

&\multicolumn{4}{c}{ALP linear \ZZ}& \multicolumn{4}{c}{ALP chiral \ZH} \\
Boosted  & \multicolumn{4}{c}{ 8} & \multicolumn{4}{c}{ 7} \\
Resolved & \multicolumn{4}{c}{ 7} & \multicolumn{4}{c}{10} \\
Combined & \multicolumn{4}{c}{15} & \multicolumn{4}{c}{17} \\
\end{tabular} 
}
\end{table*} 

Results are extracted from a combined maximum likelihood fit of signal and background to the \mZV or \mZH distribution in the 400--3000\GeV interval, simultaneously in all the categories used in the analysis.
The systematic uncertainties discussed in Section~\ref{sec:sys} are included as nuisance parameters in the maximum likelihood fit, and the background-only hypothesis is tested against the combined background and signal hypothesis~\cite{Read:2002hq,Junk:1999kv}.

No statistically significant excess is observed with respect to the SM expectations. The lowest single-sided $p$-values are of the order of 2.0\%, corresponding to a local significance at the level of 2 standard deviations, and occur for bulk gravitons with masses in the vicinity of 575, 775 and 1350\GeV, and for a $\PWpr$ boson with a mass in the vicinity of 775\GeV.
Upper limits at 95\% confidence level (\CL) on the signal production cross sections are set using the modified frequentist (\CLs) approach in the asymptotic approximation~\cite{Junk:1999kv,Read:2002hq,Cowan:2010js,CMS-NOTE-2011-005}.

\begin{figure}[htbp]
\centering
\includegraphics[width=0.49\textwidth]{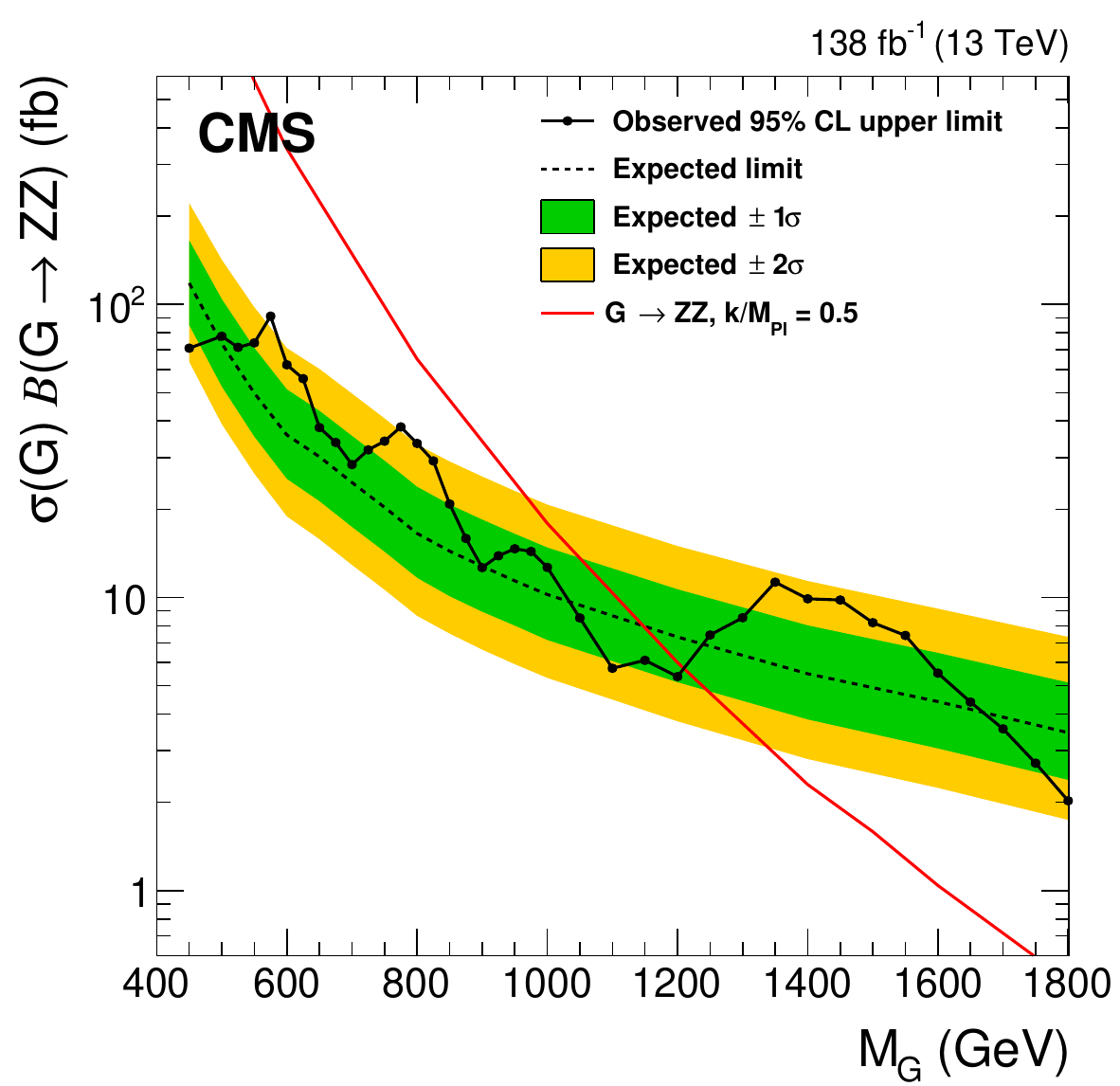}
\includegraphics[width=0.49\textwidth]{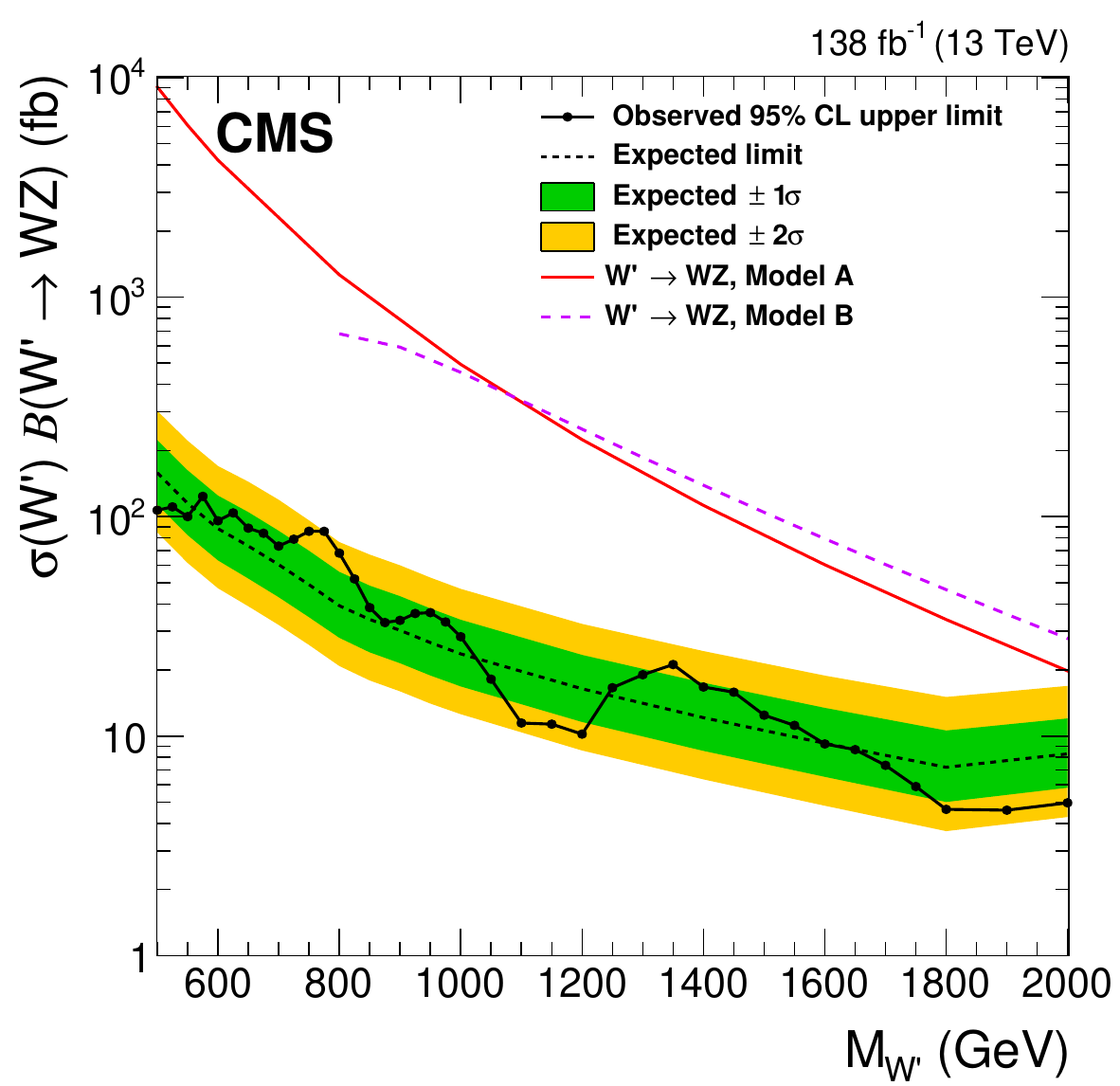}
\caption{Observed and expected 95\% \CL upper limits on $\sigma_{\PXXG} \, \B(\PXXG\to\ZZ)$ (left) and $\sigma_{\PWpr} \, \B(\PWpr\to\ZW)$ (right) as a function of the resonance mass, taking into account all statistical and systematic uncertainties. 
The electron and muon channels and the various categories used in the analysis are combined together. 
The green (inner) and yellow (outer) bands represent the 68 and 95\% coverage of the expected limit in the background-only hypothesis. 
Theoretical predictions for the signal production cross section are also shown: (left) {\PXXG} produced in the WED bulk graviton model with $\ktilde=0.5$;
(right) \PWpr produced in the framework of HVT model A and B with $g_{\PV}=1$ and 3, respectively.
}
\label{fig:limit}
\end{figure}

The observed and expected upper limits on the resonance cross section, multiplied by the branching fraction for the decay into one \PZ boson and a \PW or \PZ boson, $\sigma_{\PXXG} \, \B(\PXXG\to\ZZ)$ or $\sigma_{\PWpr} \, \B(\PWpr\to\ZW)$, are reported as a function of the resonance mass in Fig.~\ref{fig:limit}, assuming a \PWpr or {\PXXG} produced in the narrow-width approximation. A WED bulk graviton is excluded up to masses of 1200 (expected 1150)\GeV for $\ktilde = 0.5$.
This mass limit is similar to the one obtained by the CMS Collaboration in Ref.~\cite{CMS:2021itu}, and the present analysis also sets cross section limits on gravitons with a mass down to 450\GeV.
The strongest \PWpr mass exclusion limits in HVT models A and B are those quoted in Refs.~\cite{CMS:2021klu,CMS:2021itu,Aad:2020ddw}.

\begin{figure*}[htbp]
\centering
\includegraphics[width=0.49\textwidth]{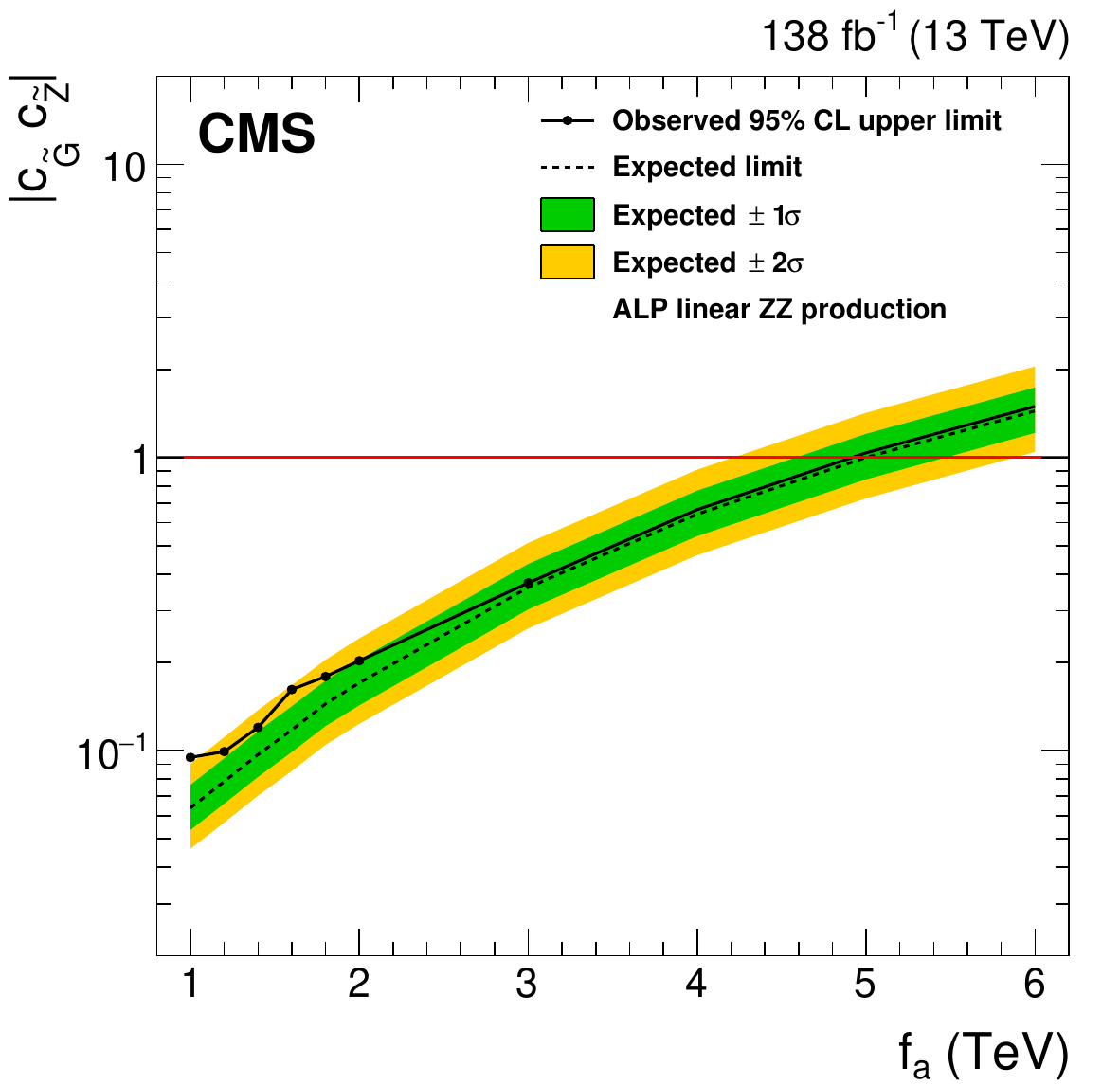} 
\includegraphics[width=0.49\textwidth]{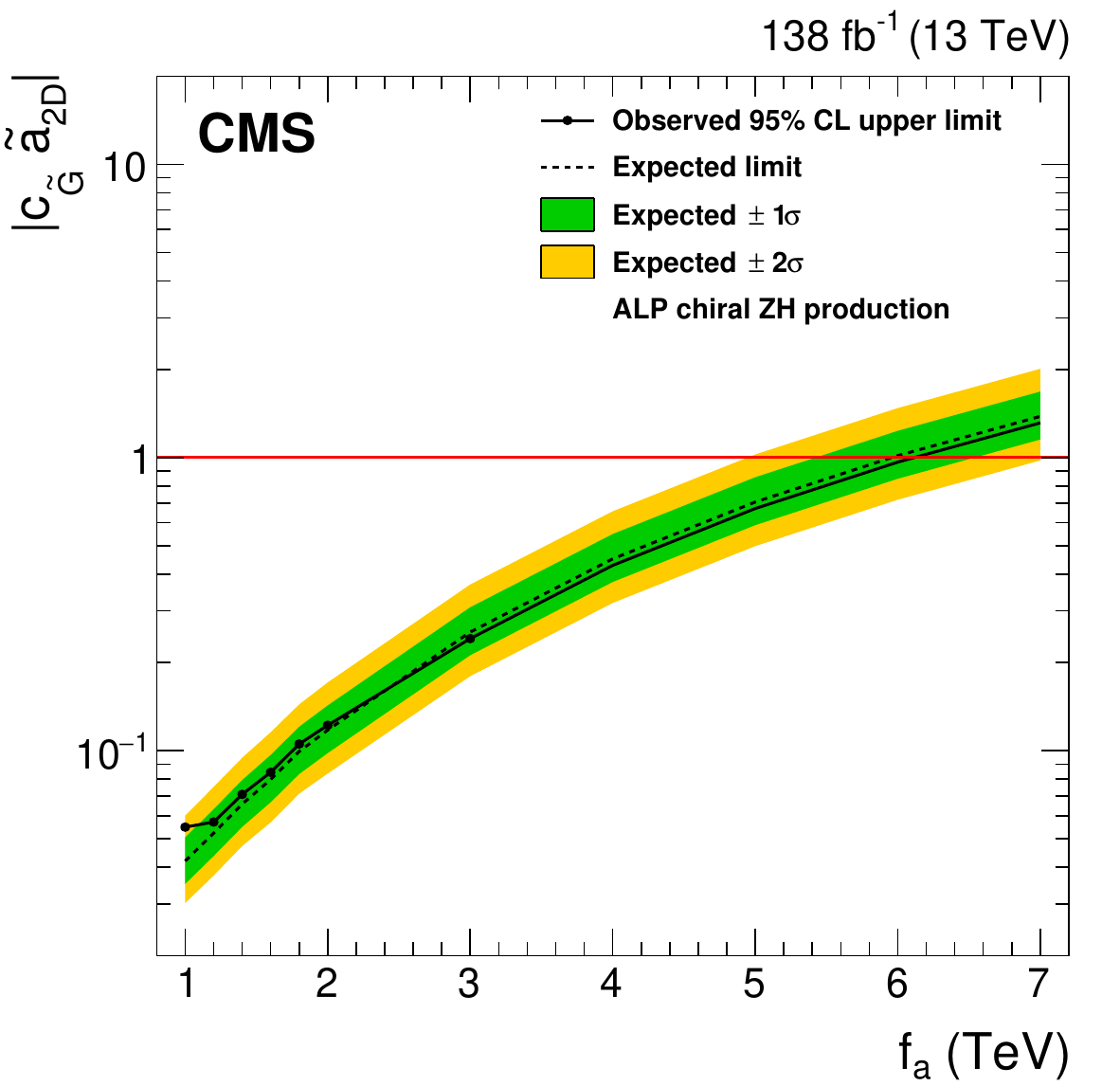}
\caption{Observed and expected 95\% \CL upper limits on the ALP linear $\abs{\cgtil \cztil}$
(left) and the ALP chiral $\abs{\cgtil \atwod}$ (right)
coupling coefficients as a function of the mass scale \falp for ALP masses $m_\alp < 100\GeV$.}
\label{fig:limit_ALP}
\end{figure*}

Figure~\ref{fig:limit_ALP} shows the observed and expected 95\% \CL upper limits on the ALP linear
$\abs{\cgtil \cztil}$ (left) and the ALP chiral $\abs{\cgtil \atwod}$ (right)
coupling coefficients, as functions of the mass scale \falp.
For each black dot in Fig.~\ref{fig:limit_ALP}, events with \mZV or $\mZH > \falp$ have been excluded from the fit, restricting the analysis to the EFT mass consistency region. Limits on the coupling coefficients for $\falp > 3\TeV$ are calculated from the same fit result as $\falp = 3\TeV$.
We define the ALP mass range where our results are considered valid as the region
where the cross section diverges by less than 10\% from its asymptotic value, $m_\alp < 100\GeV$.
The expected and observed upper limits on $\sigma(\Pg\Pg \to \astar \to \ZZ / \ZH)$ for $\falp = 3\TeV$ are reported in Table~\ref{table-limALP}.

\begin{table*}[htb]
\centering
\topcaption{Expected and observed 95\% \CL upper limits on $\sigma(\Pg\Pg \to \astar \to \ZZ / \ZH)$ in fb for $\falp = 3 \TeV$. The $\pm1\sigma$ and $\pm2\sigma$ numbers represent the 68 and 95\% coverage of the expected limit for the background-only hypothesis.}
\label{table-limALP}
\centering
\begin{tabular}{ccccccc} 
\multirow{2}{*}{Model} & \multicolumn{5}{c}{Expected} & \multirow{2}{*}{Observed} \\
 & $-2\sigma$ & $-1\sigma$ & Median & $+1\sigma$ & $+2\sigma$ & \\ \hline
ALP linear ZZ & 79 & 107 & 151 & 218 & 304 & 162 \\
ALP chiral ZH & 32 &  39 &  64 &  94 & 134 &  57 \\   
\end{tabular}
\end{table*} 

In the ALP linear model with $\falp \geq 3\TeV$, the observed (expected) 95\% \CL limit on
$\abs{\cgtil \cztil}/\falp^2$ is 0.0415 (0.0400)\TeV$^{-2}$.
The limit on $\abs{\cztil}/\falp$ and $\abs{\cgtil}/\falp$ in the case where the two are equal is 0.204 (0.200)\TeV$^{-1}$;
for $\abs{\cgtil}/\falp = 0.25\TeV^{-1}$, the limit on $\abs{\cztil}/\falp$ is 0.166 (0.160)\TeV$^{-1}$.

In the ALP chiral model with $\falp \geq 3\TeV$, the observed (expected) 95\% \CL limit on
$\abs{\cgtil \atwod}/\falp^2$ is 0.0269 (0.0281)\TeV$^{-2}$.
The limit on $\abs{\atwod}/\falp$ and $\abs{\cgtil}/\falp$ in the case where the two are equal is 0.164 (0.168)\TeV$^{-1}$;
for $\abs{\cgtil}/\falp = 0.25\TeV^{-1}$, the limit on $\abs{\atwod}/\falp$ is 0.108 (0.113)\TeV$^{-1}$.
These limits are valid for a Higgs boson with SM decays. The experimental upper limit on $\B (\PH \to \text{BSM})$ is 0.34 at 95\% \CL~\cite{ATLAS:2016neq}; thus, relaxing the SM condition would increase the limits on $\abs{\cgtil \atwod}/\falp^2$ by a factor of 1.23 at most.

\section{Summary}\label{sec:conclusions}

  A search has been presented for heavy resonances decaying to \ZZ or \ZW, and nonresonant \ZZ or \ZH production (where
\PH is the Higgs boson) mediated by axion-like particles (ALPs).
  The analysis is sensitive to resonances with masses in the range from 450 to 2000\GeV.
  Two categories are defined based on the merged or resolved reconstruction of the hadronically decaying boson.
  The search is based on data collected in 2016--2018 by the CMS experiment at the LHC in proton-proton collisions with a center-of-mass energy of 13\TeV, corresponding to an integrated luminosity of 138\fbinv.

  No significant excess is observed in the data above the standard model expectations.
Depending on the resonance mass, upper limits of 2--90 and 5--120\unit{fb}
have been set on the product of the cross section of a spin-2 bulk graviton
and the \ZZ branching fraction, and on a spin-1 \PWpr signal and the \ZW
branching fraction, respectively. 
Upper limits on the nonresonant ALP-mediated \ZZ and \ZH production cross sections for a new physics energy scale $\falp = 3\TeV$ and ALP masses $m_\alp < 100\GeV$
have been established at 162 and 57\unit{fb}, respectively.
Depending on the value of the scale \falp, upper limits on the product of the ALP coupling to gluons with the relevant coupling to \ZZ or \ZH of 0.02--0.09\TeV$^{-2}$ have been set, valid for ALP masses $m_\alp < 100\GeV$. These are the first limits based on nonresonant ALP-mediated \ZZ and \ZH production obtained by the LHC experiments.

\begin{acknowledgments}

  We congratulate our colleagues in the CERN accelerator departments for the excellent performance of the LHC and thank the technical and administrative staffs at CERN and at other CMS institutes for their contributions to the success of the CMS effort. In addition, we gratefully acknowledge the computing centers and personnel of the Worldwide LHC Computing Grid and other centers for delivering so effectively the computing infrastructure essential to our analyses. Finally, we acknowledge the enduring support for the construction and operation of the LHC, the CMS detector, and the supporting computing infrastructure provided by the following funding agencies: BMBWF and FWF (Austria); FNRS and FWO (Belgium); CNPq, CAPES, FAPERJ, FAPERGS, and FAPESP (Brazil); MES and BNSF (Bulgaria); CERN; CAS, MoST, and NSFC (China); MINCIENCIAS (Colombia); MSES and CSF (Croatia); RIF (Cyprus); SENESCYT (Ecuador); MoER, ERC PUT and ERDF (Estonia); Academy of Finland, MEC, and HIP (Finland); CEA and CNRS/IN2P3 (France); BMBF, DFG, and HGF (Germany); GSRI (Greece); NKFIA (Hungary); DAE and DST (India); IPM (Iran); SFI (Ireland); INFN (Italy); MSIP and NRF (Republic of Korea); MES (Latvia); LAS (Lithuania); MOE and UM (Malaysia); BUAP, CINVESTAV, CONACYT, LNS, SEP, and UASLP-FAI (Mexico); MOS (Montenegro); MBIE (New Zealand); PAEC (Pakistan); MSHE and NSC (Poland); FCT (Portugal); JINR (Dubna); MON, RosAtom, RAS, RFBR, and NRC KI (Russia); MESTD (Serbia); SEIDI, CPAN, PCTI, and FEDER (Spain); MOSTR (Sri Lanka); Swiss Funding Agencies (Switzerland); MST (Taipei); ThEPCenter, IPST, STAR, and NSTDA (Thailand); TUBITAK and TAEK (Turkey); NASU (Ukraine); STFC (United Kingdom); DOE and NSF (USA).

  \hyphenation{Rachada-pisek} Individuals have received support from the Marie-Curie program and the European Research Council and Horizon 2020 Grant, contract Nos.\ 675440, 724704, 752730, 758316, 765710, 824093, 884104, and COST Action CA16108 (European Union); the Leventis Foundation; the Alfred P.\ Sloan Foundation; the Alexander von Humboldt Foundation; the Belgian Federal Science Policy Office; the Fonds pour la Formation \`a la Recherche dans l'Industrie et dans l'Agriculture (FRIA-Belgium); the Agentschap voor Innovatie door Wetenschap en Technologie (IWT-Belgium); the F.R.S.-FNRS and FWO (Belgium) under the ``Excellence of Science -- EOS" -- be.h project n.\ 30820817; the Beijing Municipal Science \& Technology Commission, No. Z191100007219010; the Ministry of Education, Youth and Sports (MEYS) of the Czech Republic; the Deutsche Forschungsgemeinschaft (DFG), under Germany's Excellence Strategy -- EXC 2121 ``Quantum Universe" -- 390833306, and under project number 400140256 - GRK2497; the Lend\"ulet (``Momentum") Program and the J\'anos Bolyai Research Scholarship of the Hungarian Academy of Sciences, the New National Excellence Program \'UNKP, the NKFIA research grants 123842, 123959, 124845, 124850, 125105, 128713, 128786, and 129058 (Hungary); the Council of Science and Industrial Research, India; the Latvian Council of Science; the Ministry of Science and Higher Education and the National Science Center, contracts Opus 2014/15/B/ST2/03998 and 2015/19/B/ST2/02861 (Poland); the Funda\c{c}\~ao para a Ci\^encia e a Tecnologia, grant CEECIND/01334/2018 (Portugal); the National Priorities Research Program by Qatar National Research Fund; the Ministry of Science and Higher Education, projects no. 14.W03.31.0026 and no. FSWW-2020-0008, and the Russian Foundation for Basic Research, project No.19-42-703014 (Russia); the Programa Estatal de Fomento de la Investigaci{\'o}n Cient{\'i}fica y T{\'e}cnica de Excelencia Mar\'{\i}a de Maeztu, grant MDM-2015-0509 and the Programa Severo Ochoa del Principado de Asturias; the Stavros Niarchos Foundation (Greece); the Rachadapisek Sompot Fund for Postdoctoral Fellowship, Chulalongkorn University and the Chulalongkorn Academic into Its 2nd Century Project Advancement Project (Thailand); the Kavli Foundation; the Nvidia Corporation; the SuperMicro Corporation; the Welch Foundation, contract C-1845; and the Weston Havens Foundation (USA).
\end{acknowledgments}

\bibliography{auto_generated}

\cleardoublepage \appendix\section{The CMS Collaboration \label{app:collab}}\begin{sloppypar}\hyphenpenalty=5000\widowpenalty=500\clubpenalty=5000\cmsinstitute{Yerevan~Physics~Institute, Yerevan, Armenia}
A.~Tumasyan
\cmsinstitute{Institut~f\"{u}r~Hochenergiephysik, Vienna, Austria}
W.~Adam\cmsorcid{0000-0001-9099-4341}, J.W.~Andrejkovic, T.~Bergauer\cmsorcid{0000-0002-5786-0293}, S.~Chatterjee\cmsorcid{0000-0003-2660-0349}, K.~Damanakis, M.~Dragicevic\cmsorcid{0000-0003-1967-6783}, A.~Escalante~Del~Valle\cmsorcid{0000-0002-9702-6359}, R.~Fr\"{u}hwirth\cmsAuthorMark{1}, M.~Jeitler\cmsAuthorMark{1}\cmsorcid{0000-0002-5141-9560}, N.~Krammer, L.~Lechner\cmsorcid{0000-0002-3065-1141}, D.~Liko, I.~Mikulec, P.~Paulitsch, F.M.~Pitters, J.~Schieck\cmsAuthorMark{1}\cmsorcid{0000-0002-1058-8093}, R.~Sch\"{o}fbeck\cmsorcid{0000-0002-2332-8784}, D.~Schwarz, S.~Templ\cmsorcid{0000-0003-3137-5692}, W.~Waltenberger\cmsorcid{0000-0002-6215-7228}, C.-E.~Wulz\cmsAuthorMark{1}\cmsorcid{0000-0001-9226-5812}
% ...
\vskip\cmsinstskip
% ...
1:~Also at TU~Wien, Wien, Austria\\
% ...
\end{sloppypar}
%%% END EDITABLE REGION %%%
% skeleton_end
\end{document}